\documentclass{aa}  

\usepackage{graphicx}
\usepackage{txfonts}
\usepackage{lipsum}
\usepackage{hyperref}
\usepackage{subcaption}         
\usepackage{lscape}             
\usepackage{placeins}           
                                
\newcommand{\hiztot}{306\ }
\newcommand{\hizknown}{219\ }
\newcommand{\hiznew}{87\ }
\newcommand{\hiznewb}{19\ }
\newcommand{\hizconf}{11\ }
\newcommand{\hizconfb}{29\ }
\newcommand{\lowztot}{247\ }
\newcommand{\lowzknown}{44\ }
\newcommand{\lowznew}{203\ }

\begin{document}

   \title{Three Hundred Quasars from the Couch:\\ A first look at high-redshift quasar discovery with SPHEREx}
   \titlerunning{High-redshift quasar discovery with SPHEREx}

   \subtitle{ }

   \author{F.~B.~Davies\inst{1}\thanks{Corresponding author: davies@mpia.de} \and
           S.~E.~I.~Bosman\inst{2,1} \and
           A.~Ganguly\inst{2} \and
           E.~Ba\~{n}ados\inst{1} \and
           S.~Belladitta\inst{1,3} \and
           D.~Stern\inst{4} \and
           J.~A.~Acevedo~Barroso\inst{4} \and
           D.~Yang\inst{5} \and
           J.~F.~Hennawi\inst{5,6} \and
           F.~Wang\inst{7} \and
           J.~Yang\inst{7} \and
           X.~Fan\inst{8}
        }
    \authorrunning{F.~B.~Davies et al.}

   \institute{Max-Planck-Institut für Astronomie, Königstuhl 17,  D-69117, Heidelberg, Germany\\
            \and Institute for Theoretical Physics, Heidelberg University, Philosophenweg 12, D–69120, Heidelberg, Germany
            \and INAF - Osservatorio di Astrofisica e Scienza dello Spazio di Bologna, Via Gobetti 93/3, I-40129 Bologna, Italy
            \and Jet Propulsion Laboratory, California Institute of Technology, 4800 Oak Grove Drive, Mail Stop 264-789, Pasadena, CA 91109, USA
            \and
            Leiden Observatory, Leiden University, Leiden 2333 CA, Netherland
            \and
            Department of Physics, Broida Hall, University of California, Santa Barbara Santa Barbara, CA 93106-9530, USA
            \and
            Department of Astronomy, University of Michigan, 1085 S. University Ave., Ann Arbor, MI 48109, USA
            \and
            Steward Observatory, University of Arizona, 933 N Cherry Avenue, Tucson, AZ 85721, USA
            }

   \date{Received March XX, 2026}

  \abstract{
  Photometric selection of luminous high-redshift ($z\gtrsim4$) quasars is plagued by contamination from numerous low-mass Galactic stars, reddened lower-redshift quasars, as well as compact luminous red galaxies. Confirmation of these rare objects thus requires extensive spectroscopic campaigns on 4 and 8-meter-class telescopes with relatively low success rates. Here we demonstrate the utility of SPHEREx spectrophotometric survey data for quasar confirmation with no ground-based follow-up required, ``from the couch,'' applied to candidates from a purposefully simplistic photometric and astrometric \emph{Gaia}+\emph{WISE} selection down to low Galactic latitudes ($|b|\geq8^\circ$). Primarily from the detection of their strong broad H$\alpha$ emission lines, we discover \hiznew new luminous $4.0 < z < 5.7$ quasars with median $M_\text{1450} = -27.5$, including \hiznewb quasars at $z>5$, and recover \hizknown previously published quasars at $z>4$. We validate our SPHEREx selection with a 100\% confirmation rate in ground-based spectroscopic follow-up of \hizconfb of our new $z>4$ quasars, including \hizconf unpublished archival spectra. We also discover \lowznew additional lower-redshift quasars at $0.3 < z < 4$, consisting primarily of relatively rare highly-reddened and strong broad-absorption-line objects that are likely missed by traditional quasar surveys. Finally, we show that the Ly$\alpha$ absorption breaks and H$\alpha$ lines of luminous quasars are already detectable at redshifts $5.7\lesssim z\lesssim6.5$ after the completion of only the first of four all-sky surveys to be performed by SPHEREx during its planned two-year mission.
  }

   \keywords{quasars: general, Astronomical data bases }

   \maketitle

    \nolinenumbers

\section{Introduction}

\begin{figure*}
    \centering
    \includegraphics[width=0.9\linewidth]{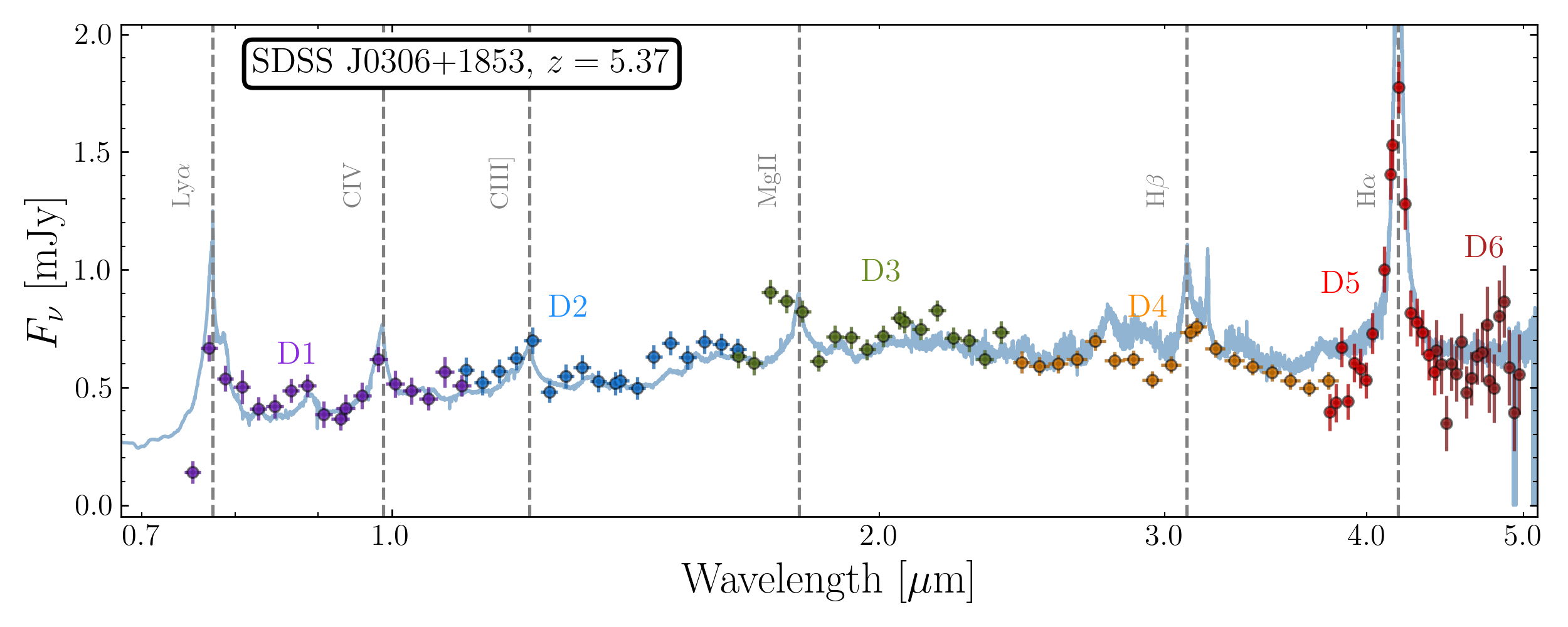}
    \vskip-1em
    \caption{SPHEREx spectrophotometry of the luminous ($M_{1450} = -28.9$) previously-known quasar SDSS J0306$+$1853 \citep{Wang15} at $z=5.37$ \citep{Brazzini25} is shown by the colored points, and the blue curve shows the \citet{Selsing16} quasar template for comparison. The labels D1 through D6 indicate the six detectors of SPHEREx probing consecutive spectrophotometric bands. The most prominent features enabling the quasar's idenfication in SPHEREx data are the broad and strong H$\alpha$ line and the Ly$\alpha$ continuum break.}
    \label{fig:j0306}
\end{figure*}

\begin{figure}
    \centering
    \includegraphics[width=1.0\linewidth]{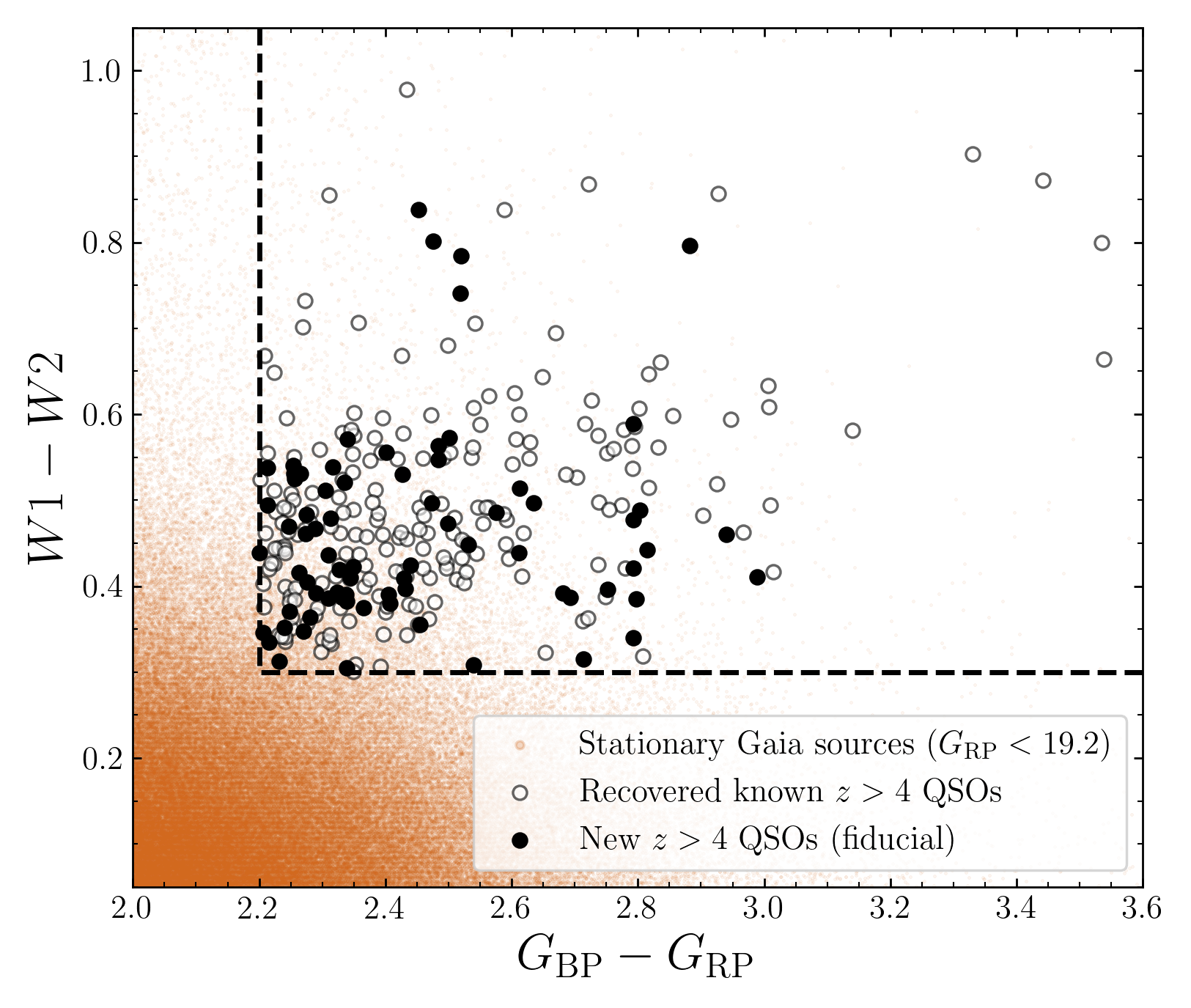}
    \caption{Illustration of our fiducial target selection in color space. The initial parent sample selected from \emph{Gaia} photometry is shown by the cloud of small brown points. Sources lying above and to the right of the dashed line were chosen for extraction of SPHEREx photometry. The black points show the newly discovered quasars (filled circles) and known quasars recovered by our search (open circles).}
    \label{fig:w1w2_rp}
\end{figure}

The most luminous quasars in the Universe probe the most massive and fastest accreting supermassive black holes (SMBHs) across cosmic time.
Characterization of their population via the luminosity distribution function across cosmic time allows for the detailed study of SMBH growth demographics as well as SMBH-galaxy co-evolution (e.g. \citealt{Soltan82,Lapi06,Silverman08,Marulli08,Giallongo15,Kulkarni19qlf}). Due to their large luminosities which enable follow-up observations with relative ease, samples of the brightest quasars with high redshifts $z\gtrsim4$ are of sweeping use across astrophysics, yielding pioneering results on cosmic metal enrichment (e.g.~\citealt{Rafelski12,Becker15,RDavies23}), SMBH growth processes (e.g.~\citealt{Volonteri10,Inayoshi20}), large-scale structure formation and the intergalactic medium (e.g.~\citealt{HM12,Boera19,Becker21}), fundamental physics (e.g.~\citealt{Martins17,Wilczynska20}) and cosmic reionization (e.g.~\citealt{Fan06,Davies18b,Bosman22}), to name but a few. The efficient discovery of new distant luminous quasars is therefore an area of active inquiry (e.g.~\citealt{Banados16,Nanni22,Cristiani23,Belladitta25}).

Identification of these extreme objects, particularly at $z>5$, is challenging: they have optical broadband magnitudes and colors similar to late-type stars in the Milky Way, yet they are more than a thousand times less numerous. Sophisticated selection techniques, including the use of infrared photometry, precise astrometric measurements, and machine-learning methods, have nevertheless resulted in samples of hundreds of such quasars following exhaustive optical spectroscopic follow-up campaigns \citep{fan23}. Sufficiently deep all-sky spectroscopy would negate the need for such follow-up, but the completeness of large surveys such as the Sloan Digital Sky Survey (SDSS; \citealt{Lyke20}) and the Dark Energy Spectroscopic Instrument survey (DESI; \citealt{DESI_qso}) at their relatively faint magnitudes ($m_{\rm AB}\lesssim19.5$) is still low. The \textit{Euclid} mission has recently demonstrated the potential of space-based slitless spectroscopy to identify high-redshift quasars, confirming sources up to $z\sim5.5$ \citep{Banados25,Fu26}. In this paper, we demonstrate the potential for quasar discovery of The SpectroPhotometer for the History of the Universe, Epoch of Reionization, and Ices eXplorer satellite (SPHEREx; \citealt{SPHEREx,Bock25}).

SPHEREx was launched on the 11$^{\rm{th}}$ of March 2025, and represents a significant leap in our knowledge of the near-infrared sky. The data consist of images taken using a Linear Variable Filter (LVF), such that any given point on the detector is sensitive to a narrow wavelength bandpass which smoothly varies along one axis of the image. The LVF positioned in front of each of the six SPHEREx detectors is sensitive to a different range of wavelengths, corresponding to six ``bands'', D1--D6, providing a continuous wavelength coverage from $0.75$ to $5.0$\,$\mu$m. The effective spectral resolution is given by the width of the LVF bandpass, which ranges from $R\sim35-39$ for D1$-$D4 ($0.75\,\mu{\rm m} < \lambda < 3.8\,\mu{\rm m}$ to $R\sim120$ for D5 and D6 ($3.8\,\mu{\rm m} < \lambda < 5.0\,\mu{\rm m}$. The survey performs 17 short steps along the wavelength axis of the LVFs for complete spectral coverage. Each point on the sky is imaged in this way on every detector, resulting in a typical $17\times6=102$ spectral channels per all-sky survey.

Scaling from the full survey depth forecast in \citet{Bock25} to include four all-sky passes, the $5\sigma$ depth of the first all-sky survey, recently complete at the time of writing, is just shy of 19 AB mag across most of the SPHEREx bandpass. Quasars of this apparent magnitude or brighter should thus already be well-characterized. In Figure~\ref{fig:j0306}, we show SPHEREx spectrophotometry for the most luminous quasar at $z\sim5.5$ ($M_\text{1450} = -28.9$), SDSS J0306$+$1853 \citep{Wang16}, obtained from the publicly available SPHEREx Quick Release data via the NASA/IPAC Infrared Science Archive (IRSA) Spectrophotometry Tool. The detailed spectral energy distribution (SED) of this quasar is clearly visible across the entire SPHEREx bandpass, which includes a first look at the spectrum at wavelengths longer than $\sim2.5$\,$\mu$m. The broad H$\alpha$ 6564\AA \ line of the quasar at $\lambda_{\rm obs}=4.18$\,$\mu$m is especially prominent, and partially spectroscopically resolved by the $R\sim120$ resolution in the D5 SPHEREx band. We leave quantitative analysis of this spectrum to future work; we highlight it here to demonstrate that the spectra of luminous quasars can already be characterized with the first all-sky survey of SPHEREx. The strength of the H$\alpha$ line detection, along with the Ly$\alpha$ forest break and other weaker emission lines, motivates the use of SPHEREx survey data as a luminous high-redshift quasar confirmation tool (e.g.~\citealt{Feder24}).

\begin{figure*}
    \centering
    \includegraphics[width=0.85\linewidth]{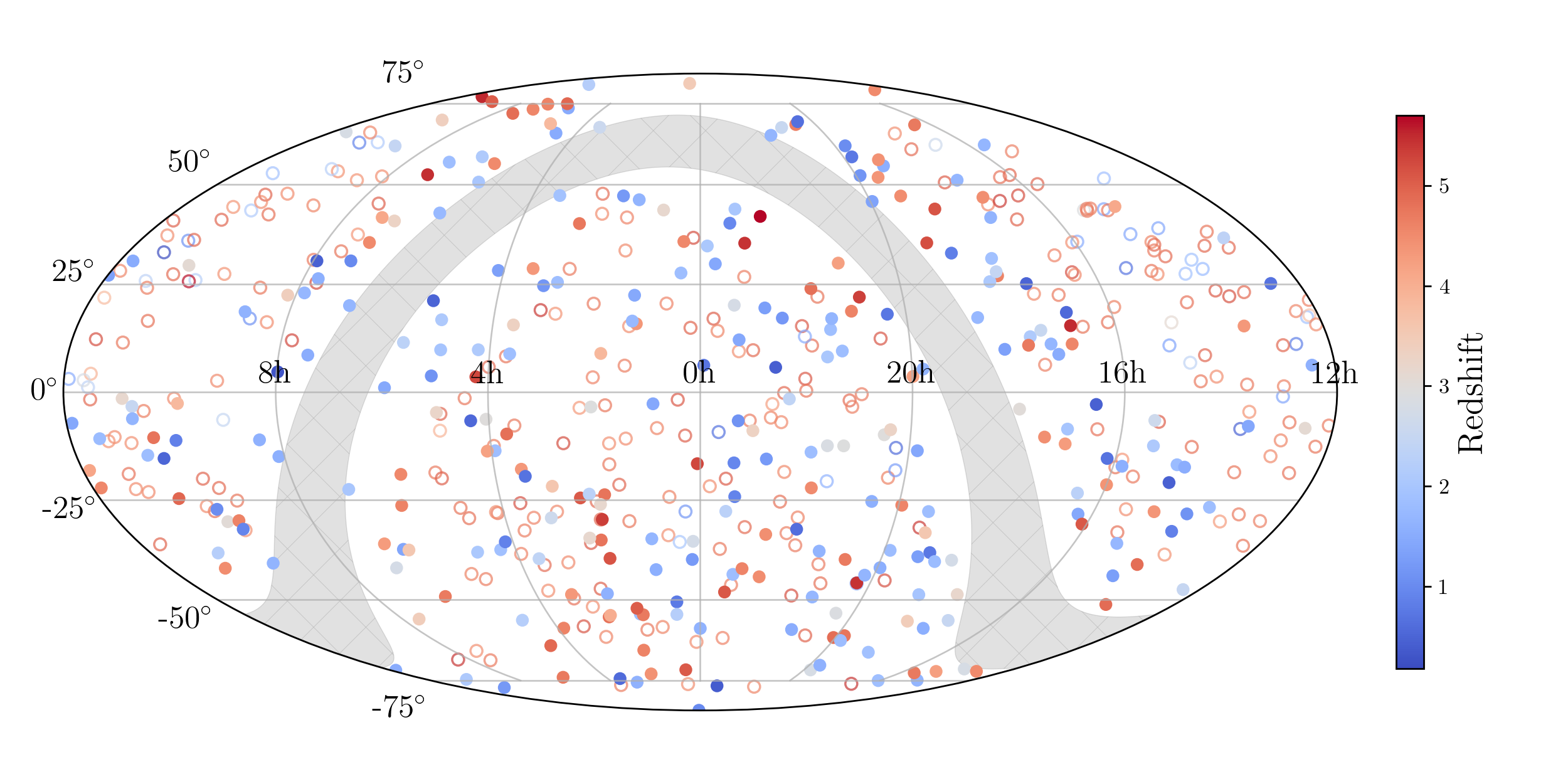}
    \vskip -1em
    \caption{Mollweide projection map of new quasars (solid circles) and recovered known quasars (open circles) identified with SPHEREx spectrophotometry. The exclusion region above and below the Galactic plane ($|b|\leq8^\circ$) is shown with a grey band.}
    \label{fig:map}
\end{figure*}

In this work, we use SPHEREx spectrophotometry to search for luminous quasars. We perform a naive all-sky selection based solely on \emph{Gaia} astrometry \citep{Gaia}, \emph{Gaia} $G_{\rm BP}-G_{\rm RP}$ photometric color, and $W1-W2$ color from the Wide-field Infrared Survey Explorer (\emph{WISE}; \citealt{WISE,NeoWISE}) as detailed in Section~{\ref{sec:phot}}. In Section~{\ref{sec:ID}}, we then ``follow up'' the resulting thousands of candidates by extracting their SPHEREx spectra using the IRSA Spectrophotometry Tool. From this large candidate pool, we identify a new sample of luminous quasars primarily via their broad H$\alpha$ emission lines. Finally, we demonstrate that even the first all-sky survey of SPHEREx is capable of detecting prominent spectral features of even higher redshift quasars, with H$\alpha$ and Ly$\alpha$ breaks clearly visible in known quasars up to $z\simeq6.5$. We discuss future prospects and conclude in Section~\ref{sec:future} and Section~\ref{sec:ccl}, respectively.

For conversion of apparent to absolute magnitudes, we assume a flat $\Lambda$CDM cosmology with $h=0.6766$ and $\Omega_m=0.3097$ \citep{Planck18}. Apparent magnitudes are given relative to Vega and absolute magnitudes are given in the AB system \citep{Oke83}, unless specified otherwise.

\section{Photometric selection}
\label{sec:phot}

High-redshift quasars are characterized by a roughly power-law UV-to-optical spectrum with a strong break due to the onset of Ly$\alpha$ forest absorption at $\lambda_{\rm rest} < 1215.67$\,\AA. Our goal in this work is not to pioneer a new, efficient selection method; rather, we make use of the inexpensive ``follow-up spectroscopy'' enabled by the SPHEREx survey to overcome low selection efficiency. By aiming for a simple, relatively inefficient selection, we hope to identify quasars which are less distinguishable from contaminants by their photometry alone. In addition, we can relax restrictions on Galactic latitude, pushing into regions of sky which have traditionally never been explored by high-redshift quasar searches due to the high likelihood of stellar contamination \citep[e.g.,][]{Yang19b,Banados23}.

We construct our selection using a combination of \emph{Gaia} astrometry and photometry, as well as \emph{WISE} photometry, inspired by \citet{Wolf20} and \citet{Onken22}. 
For our fiducial quasar candidate selection, we select relatively bright \emph{Gaia} objects ($G_{\rm RP}<19.2$\,mag) with a red optical color ($G_{\rm BP}-G_{\rm RP}>2.2$) and no detected parallax or proper motion in \emph{Gaia} Data Release 3 \citep{GaiaDR3}. Specifically, we exclude objects with a (positive) parallax detected at more than $2\sigma$, and require a total proper motion signal-to-noise less than $3$ (see also \citealt{Storey-Fisher24}). The astrometric selection will exclude the vast majority of Milky Way stars, while the red color will select for the onset of intergalactic Ly$\alpha$ and Lyman-series forest absorption at $z\gtrsim4.3$ (see \citealt{Yang17,Yang19b}). To remove optically-red stellar contaminants, we refine the selection by requiring a red color also in the rest-frame infrared ($W1-W2>0.3$) measured by \emph{WISE}, via the CatWISE2020 catalog \citep{CatWISE2020}. The final photometric selection consists of 3076 objects; we will refer to this selection as the ``fiducial'' selection.

In an attempt to push the selection to fainter and higher redshift objects, we employed two additional tuned selections. First, for the ``faint $G_{\rm RP}$'' selection, we relaxed the brightness criterion in $G_{\rm RP}$ to $<19.5$\,mag and the color criterion to $G_{\rm RP}-G_{\rm BP}>1.8$. 
Our second alternative selection, the ``$G_{\rm BP}$ flux'' selection, goes even fainter to $G_{\rm RP}<19.7$\,mag, and replaces the $G_{\rm BP}-G_{\rm RP}$ color cut with an upper limit on the $G_{\rm BP}$ signal-to-noise ratio $F(G_{\rm BP})/\sigma(G_{\rm BP})<6$, noting that known bright $z>5.5$ objects present in the \emph{Gaia} DR3 catalog are sometimes erroneously detected at this level due to the peculiarities of the \emph{Gaia} DR3 data processing \citep{Riello21}. To ensure low absolute values of $G_{\rm BP}$ flux we also required $\sigma(G_{\rm BP})<10$\,e$^{-}$\,s$^{-1}$. For both of these additional selections, we further restrict the range of WISE colors to $0.65 < W1-W2 < 1.0$ to reduce contamination from both Galactic stars and $z<4$ quasars, and also restricted the search to slightly larger Galactic latitudes $|b|\geq10^\circ$.. The $G_{\rm RP}$ faint selection resulted in 1263 additional candidates, while the $G_{\rm BP}$ flux selection resulted in 546 additional candidates.

\section{Quasar identification with SPHEREx}
\label{sec:ID}

\begin{figure*}
    \centering
    \includegraphics[width=0.32\linewidth]{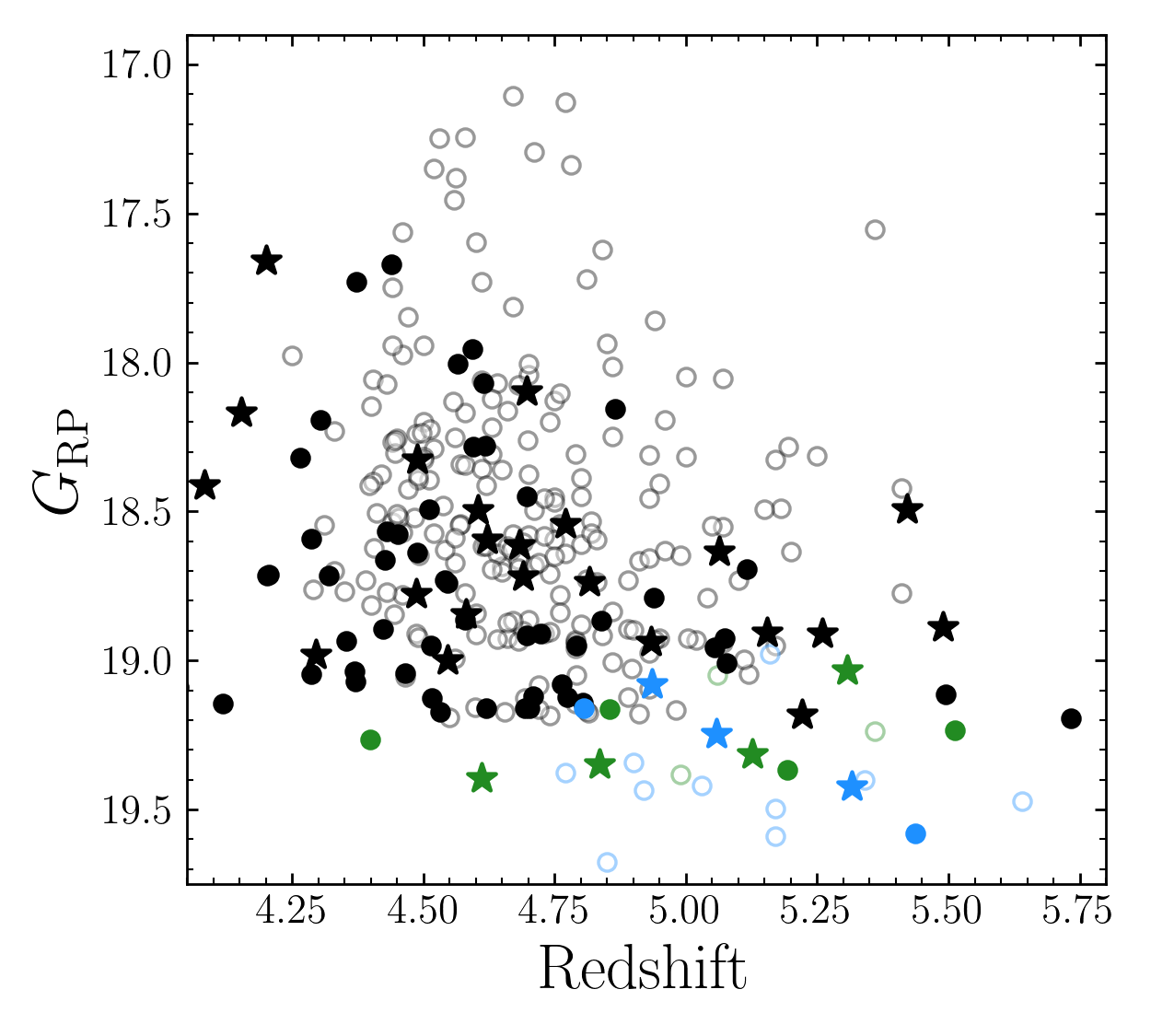}
    \includegraphics[width=0.32\linewidth]{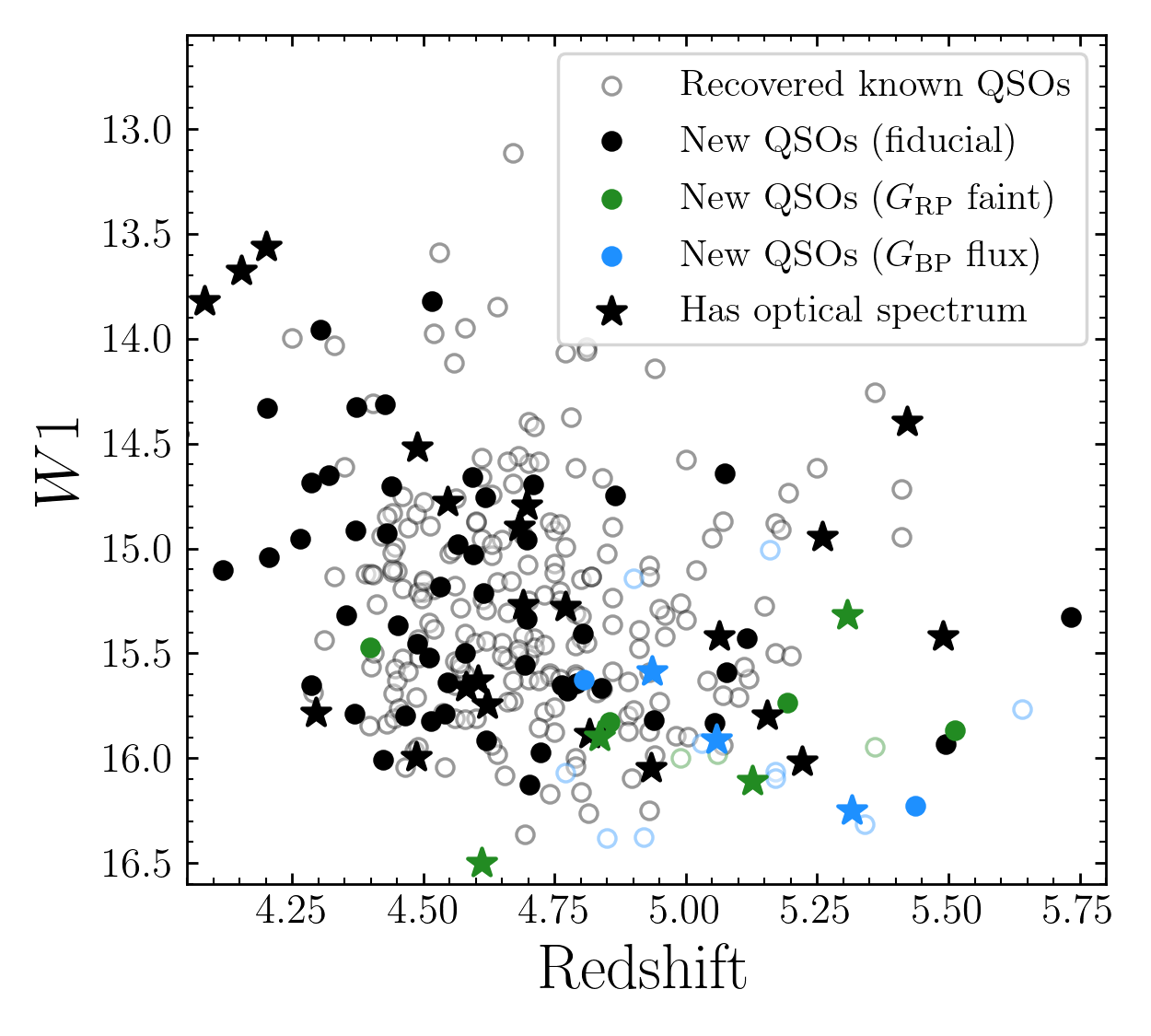}
    \includegraphics[width=0.32\linewidth]{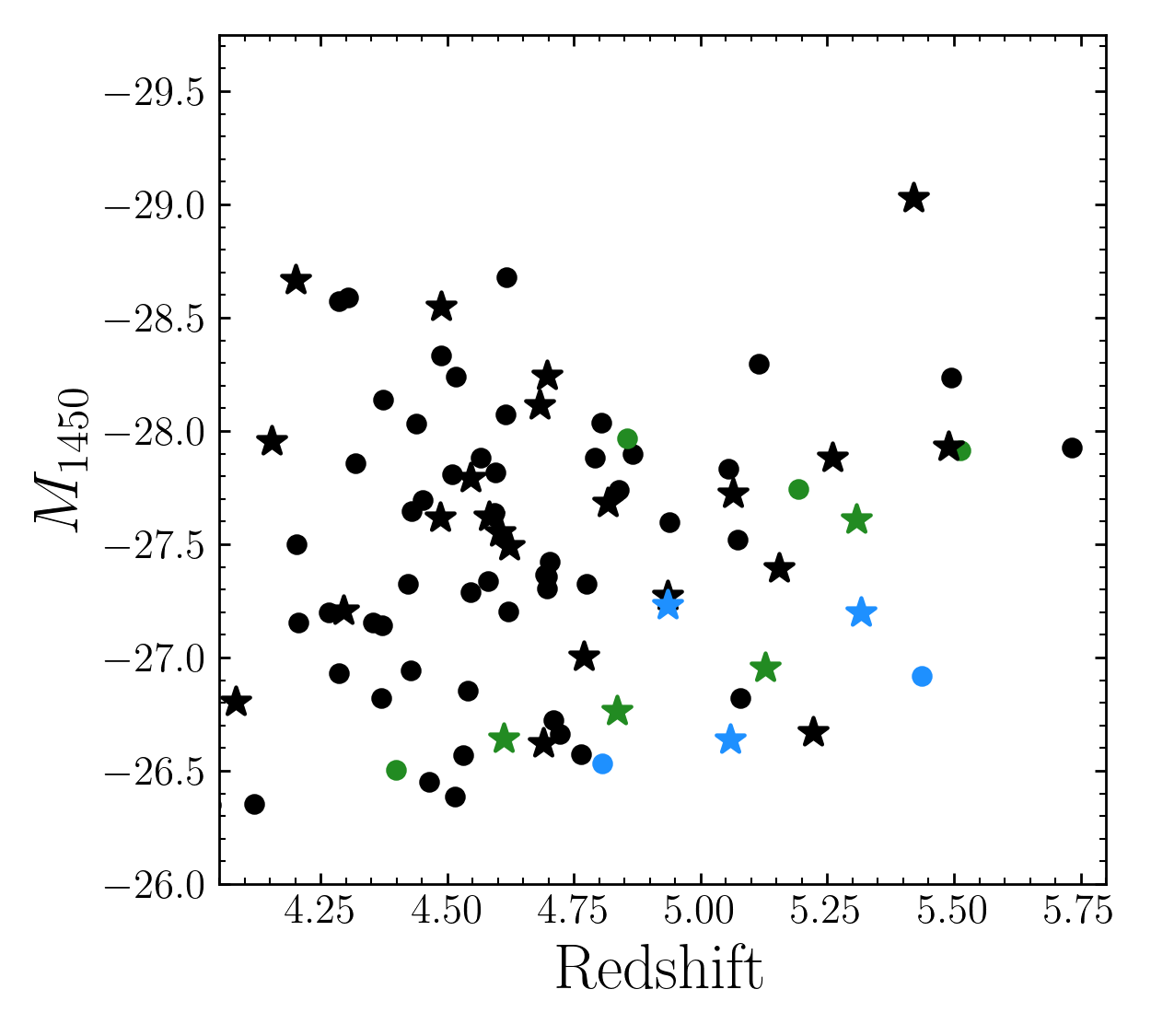}
    \vskip-0.5em
    \caption{Distribution of the $G_{\rm RP}$ (left) and $W1$ (middle) magnitudes for new (filled circles) and previously known (open circles) $z>4$ quasars in our search, as well as the $M_{1450}$ values derived from template fitting (right). Black points were found by the fiducial selection, while the green and blue points were found by the $G_{\rm RP}$ faint and $G_{\rm BP}$ flux selections, respectively. Star-shaped points denote quasars with optical confirmation spectra in the ESO archive or in this work (Section~\ref{sec:fup}).}
    \label{fig:rp_w1_z}
\end{figure*}

\begin{figure*}
    \centering
    \includegraphics[width=1.0\linewidth]{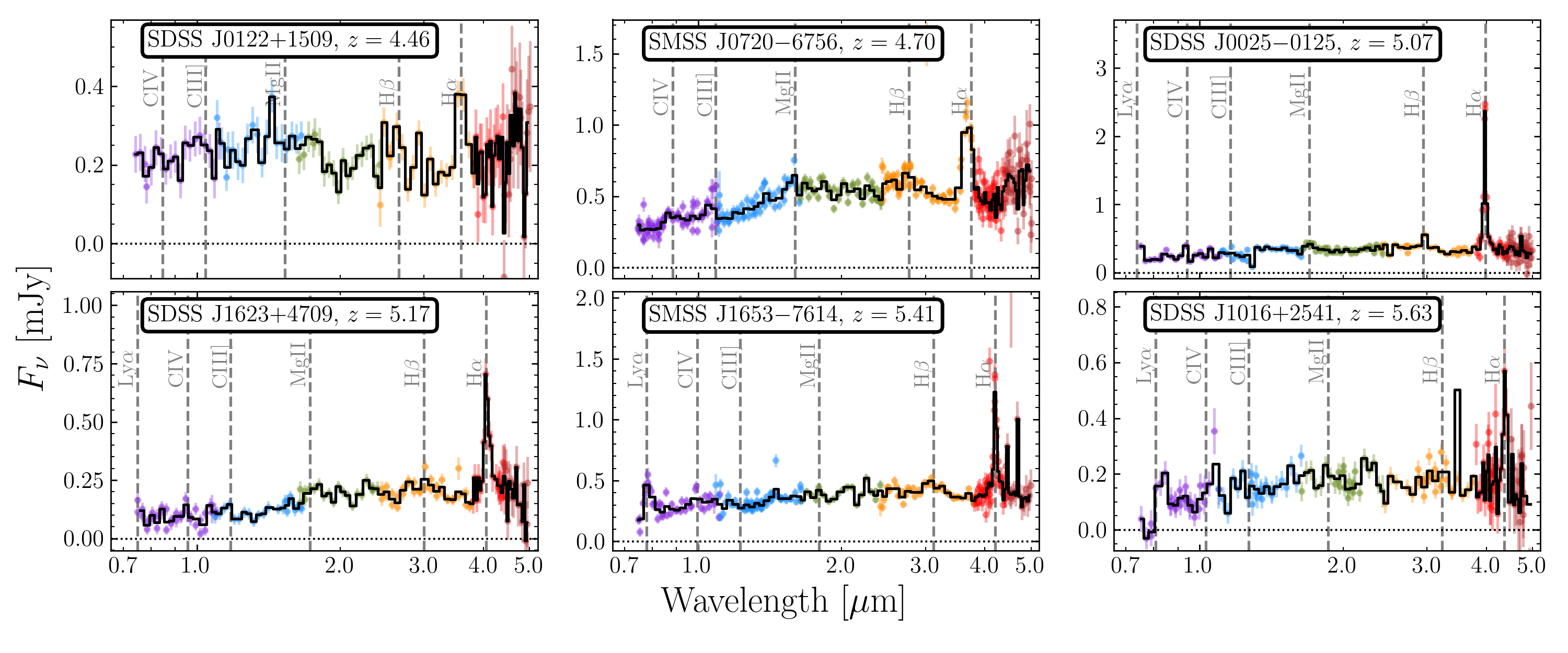}
    \vskip-1em
    \caption{Example spectrophotometry of previously known quasars recovered by our SPHEREx quasar search. The small transparent rainbow points show the raw measurements from the individual SPHEREx detectors (e.g.~Figure~\ref{fig:j0306}), 
    while the black curve is re-binned onto a common wavelength grid. 
    The wavelengths of common emission lines are shown by the vertical dashed lines.}
    \label{fig:hiz_known}
\end{figure*}

We use the SPHEREx Spectrophotometry Tool\footnote{\href{https://irsa.ipac.caltech.edu/applications/spherex/tool-spectrophotometry}{https://irsa.ipac.caltech.edu/applications/spherex/tool-spectrophotometry}} to extract low-resolution ($R\sim35$--$120$) 0.75$-$5.00\,$\mu$m spectra from the Quick Release data products of the ongoing SPHEREx survey \citep{Bock25}. In brief, the Spectrophotometry Tool performs forced PSF photometry on the LVF images using the Tractor \citep{Tractor} and a detailed model for the SPHEREx PSF as a function of wavelength and detector position. More details can be found in the SPHEREx Explanatory Supplement\footnote{\href{https://irsa.ipac.caltech.edu/data/SPHEREx/docs/SPHEREx_Expsupp_QR_v1.0.pdf}{https://irsa.ipac.caltech.edu/data/SPHEREx/docs/SPHEREx\_ \\ Expsupp\_QR\_v1.0.pdf}}. We subsequently corrected the spectrophotometry for Galactic extinction using the \citet{SB11} $E(B-V)$ map queried via IRSA and the \citet{Fitzpatrick99} extinction curve.

At the time of writing, the Spectrophotometry Tool was limited to processing two catalogs of 20 targets each at any given time. The candidates selected above were thus separated into sets of 20 objects, and individually uploaded manually to IRSA for processing. Upon completion, the spectra were retrieved and inspected by eye for quasar features, i.e.~broad emission lines of hydrogen, predominantly H$\alpha$, H$\beta$, Paschen-$\alpha$, and Paschen-$\beta$, as well as overall consistency with a quasar SED shape vs.\ the blackbody spectrum of typical stellar contaminants. All objects of interest were also visually inspected with ALADIN \citep{Bonnarel00} in the deepest available optical photometry -- from the Panoramic Survey Telescope and Rapid Response System (Pan-STARRS; \citealt{Chambers16}), DESI Legacy Surveys \citep{Dey19}, or SkyMapper \citep{Onken24} -- to qualitatively confirm their nature and check for possible contamination by nearby objects.

Due to the large $\simeq$6.2-arcsecond pixel scale of SPHEREx, and the relatively large diffraction limit of its 20\,cm aperture in the near-infrared, nearby object contamination is not so uncommon ($\lesssim10\%$ of candidates). For all candidates with interesting spectral features suggesting a high redshift ($z>4$) solution, i.e.~a strong broad emission line redward of $\simeq$3.3\,$\mu$m, we used the deblending feature of the  Spectrophotometry Tool, which performs spectral deblending by using manually uploaded coordinates of objects in CatWISE2020 within 25 arcseconds of the candidate spectrum. By default, the Spectrophotometry Tool does not include this deblending step due to high computational costs. Anecdotally, this extra step enabled a positive determination of high-redshift status for about half of the blended candidates, while the other half were either disproven or remained ambiguous.

\subsection{Spectrophotometric identifications}

We identified emission-line objects by visual inspection of all candidate spectra. Redshifts were first determined by eye from 

\newpage

\begin{figure*}
    \includegraphics[width=\linewidth]{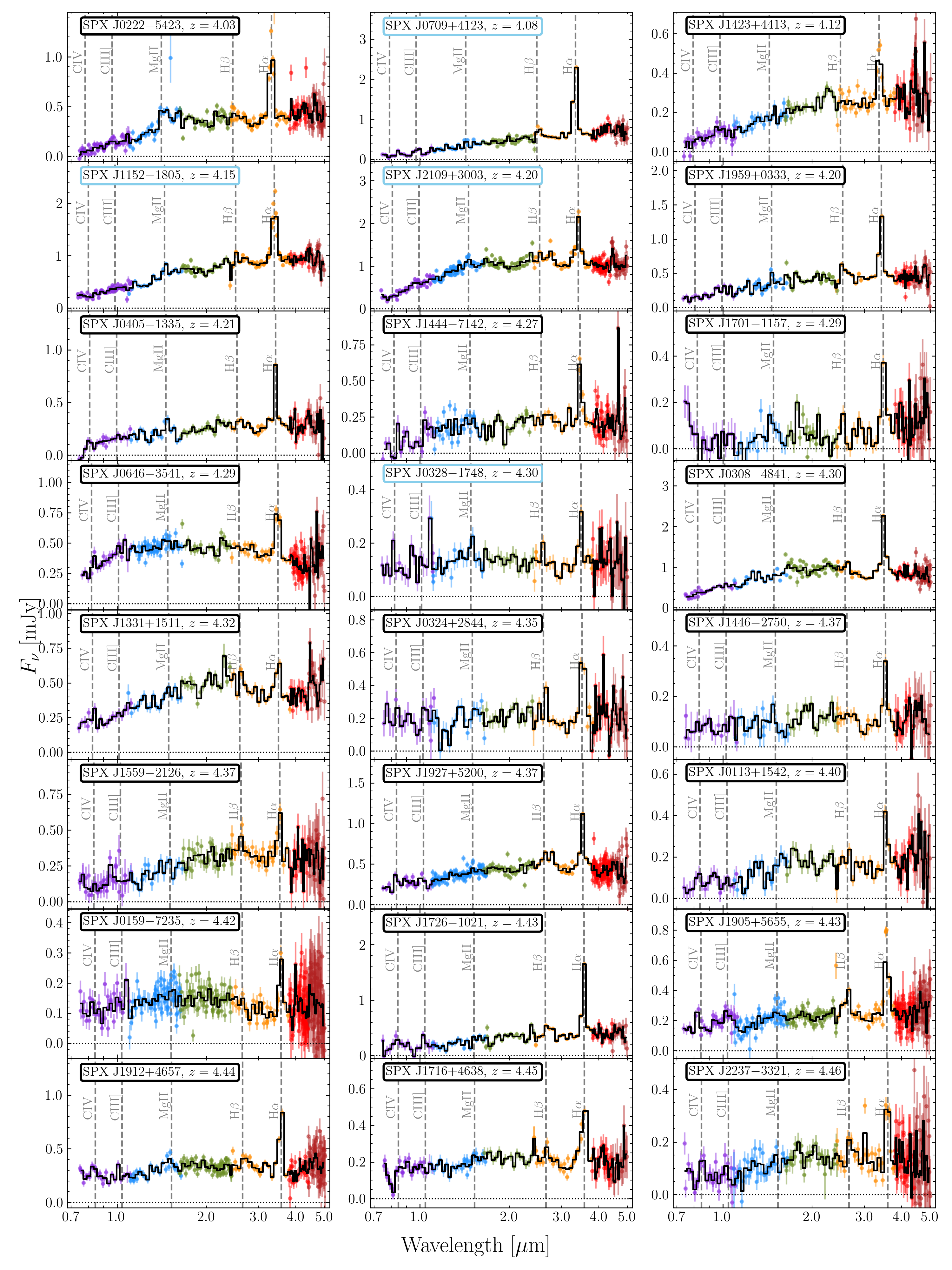}
    \caption{Gallery of new $4 < z < 5$ quasars identified in our SPHEREx search. Similar to Figure~\ref{fig:hiz_known}, the wavelengths of common emission lines are shown by the vertical dashed lines.}
    \label{fig:z45_1}
\end{figure*}

\clearpage

\begin{figure*}
    \ContinuedFloat
    \centering
    \includegraphics[width=\linewidth]{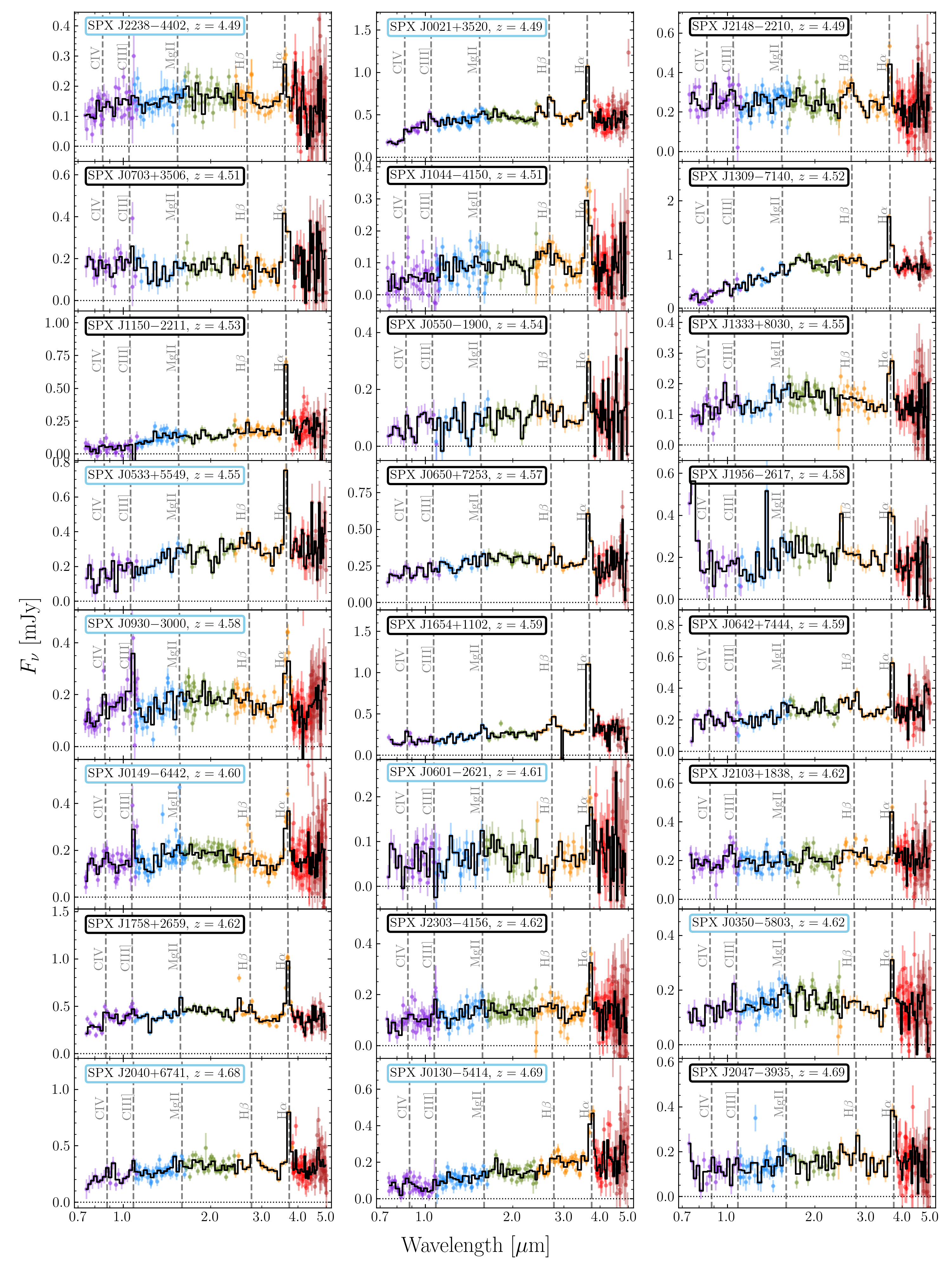}
    \caption{Continued.}
    \label{fig:z45_2}
\end{figure*}

\clearpage

\begin{figure*}
    \ContinuedFloat
    \centering
    \includegraphics[width=\linewidth]{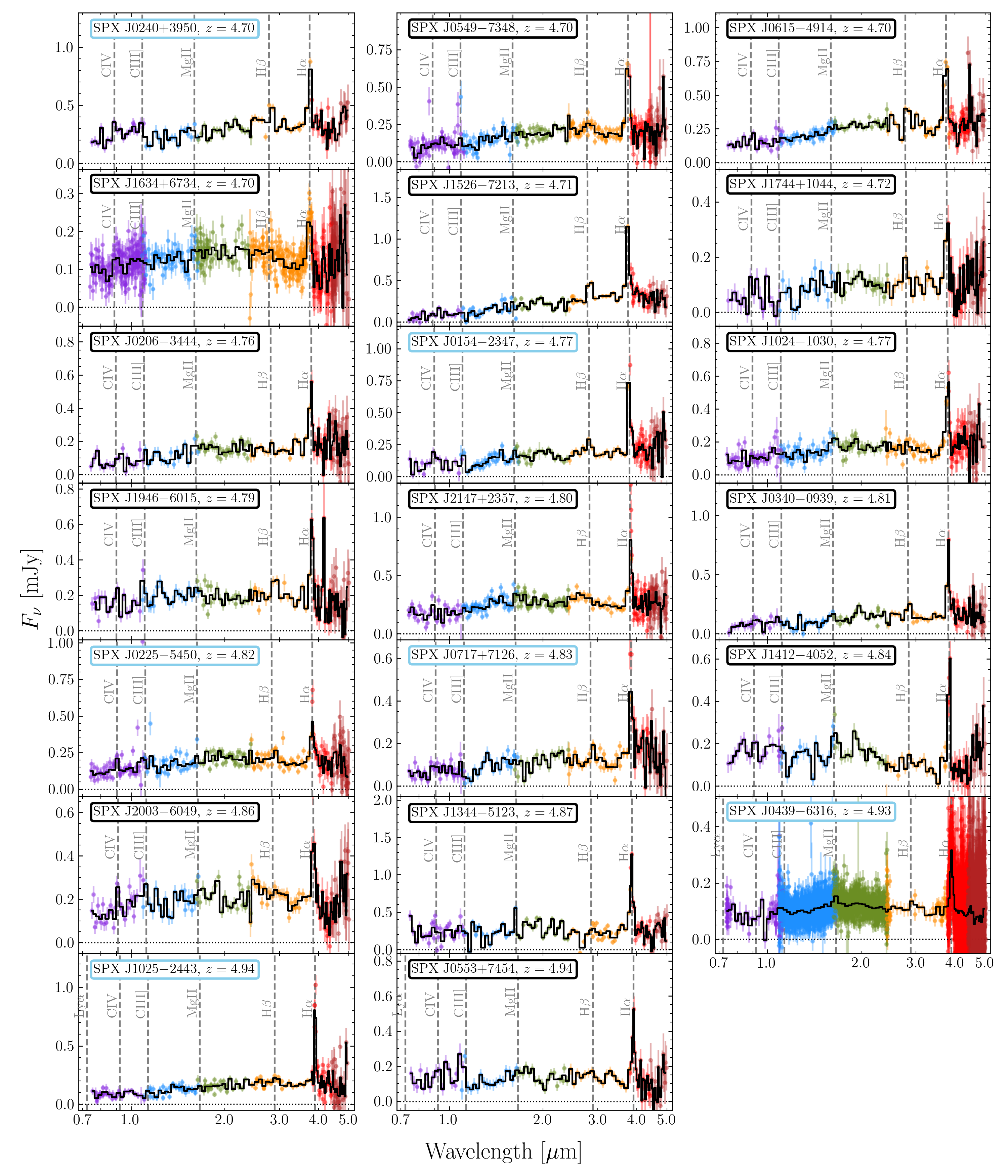}
    \caption{Continued.}
    \label{fig:z45_3}
\end{figure*}

\clearpage

\begin{figure*}
    \centering
    \includegraphics[width=\linewidth]{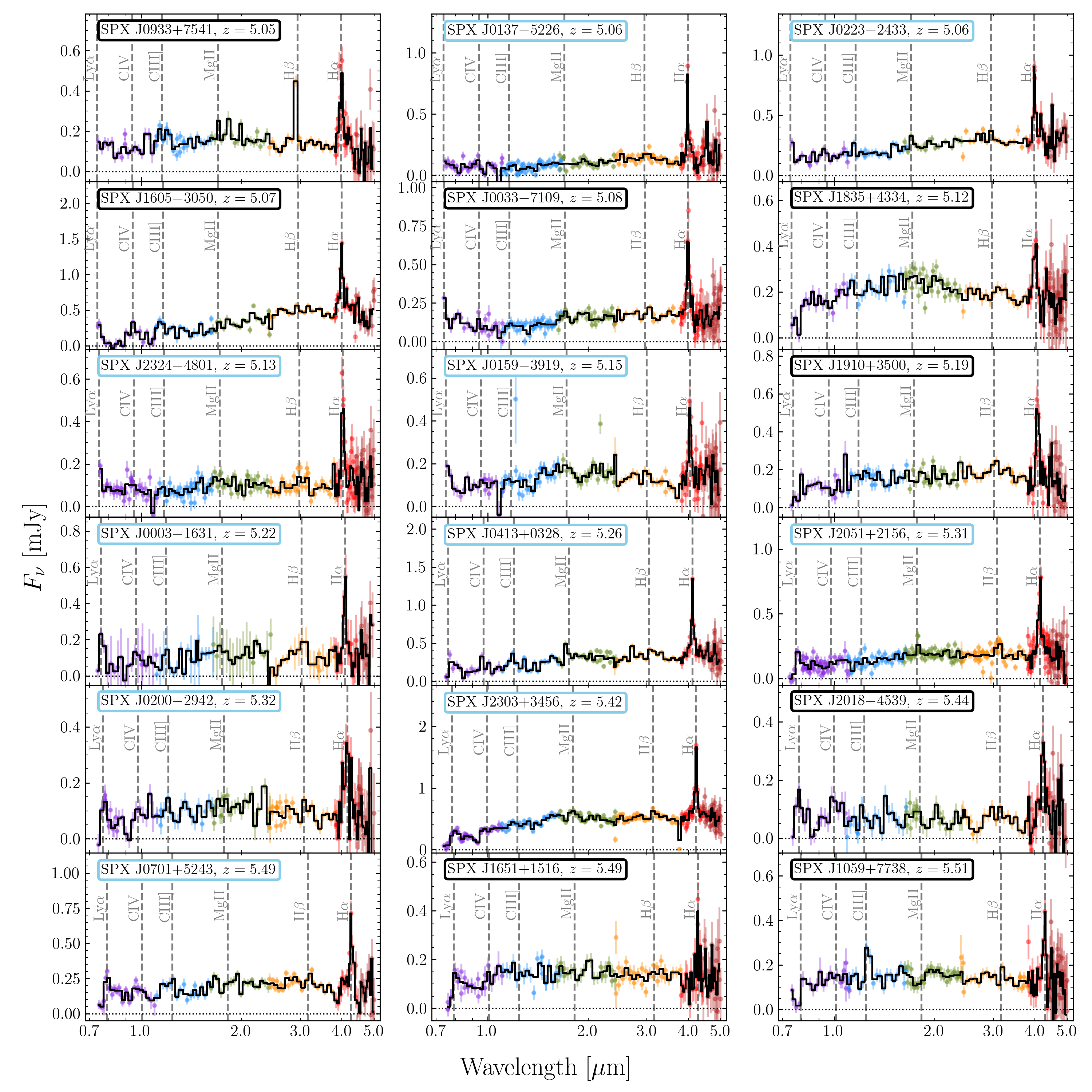}
    \caption{Similar to Figure~\ref{fig:z45_1} but for newly identified $5<z<5.7$ quasars.}
    \label{fig:z5}
\end{figure*}

\clearpage

\noindent the strongest visible emission line(s), with H$\alpha$ as the strongest line in the vast majority of cases. Our primary targets with $z>4$ subsequently had a more precise redshift determined from fitting the H$\alpha$ emission line and its surroundings with a Gaussian plus local power-law continuum model. The resulting formal precision on the redshift is typically $\Delta z \sim 0.01$,  although we note that we have not accounted for the sub-Nyquist wavelength sampling,  and the flux uncertainties from the Spectrophotometry Tool show indications of being underestimated in some cases. To approximately account for these additional uncertainties, we adopt a fixed redshift error of $\Delta z = 0.03$.

In total, we identified \hiztot quasars at $z>4$, consisting of \hizknown known quasars\footnote{Note that we did not remove or flag previously known quasars during the selection process, i.e.~the identifications were entirely ``blind'' except for a small handful of cases where the coordinates were familiar to the authors, e.g. SDSS\,J0306+1853 (Figure~\ref{fig:j0306}).} and \hiznew new quasars. The candidates' distribution across the sky is shown in Figure~\ref{fig:map}, demonstrating that many of the new bright quasars are located within 20 degrees of the Galactic plane where previous large search efforts were most incomplete. 
The corresponding distribution of $G_{\rm RP}$ and $W1$ magnitudes with redshift are shown in Figure~\ref{fig:rp_w1_z}. 
After compiling our list of new quasars, we found that at least \hizconf of them have been confirmed by other groups (mostly the QUBRICS survey, e.g.~\citealt{Guarneri22,Cristiani23}) by searching for their coordinates in the ESO archive\footnote{\href{https://archive.eso.org/cms.html}{https://archive.eso.org/cms.html}}, but whose confirmation has not been published before now. These unpublished objects are highlighted with star-shaped points in Figure~\ref{fig:rp_w1_z}. 

\begin{figure}
    \centering
    \includegraphics[width=1.0\linewidth]{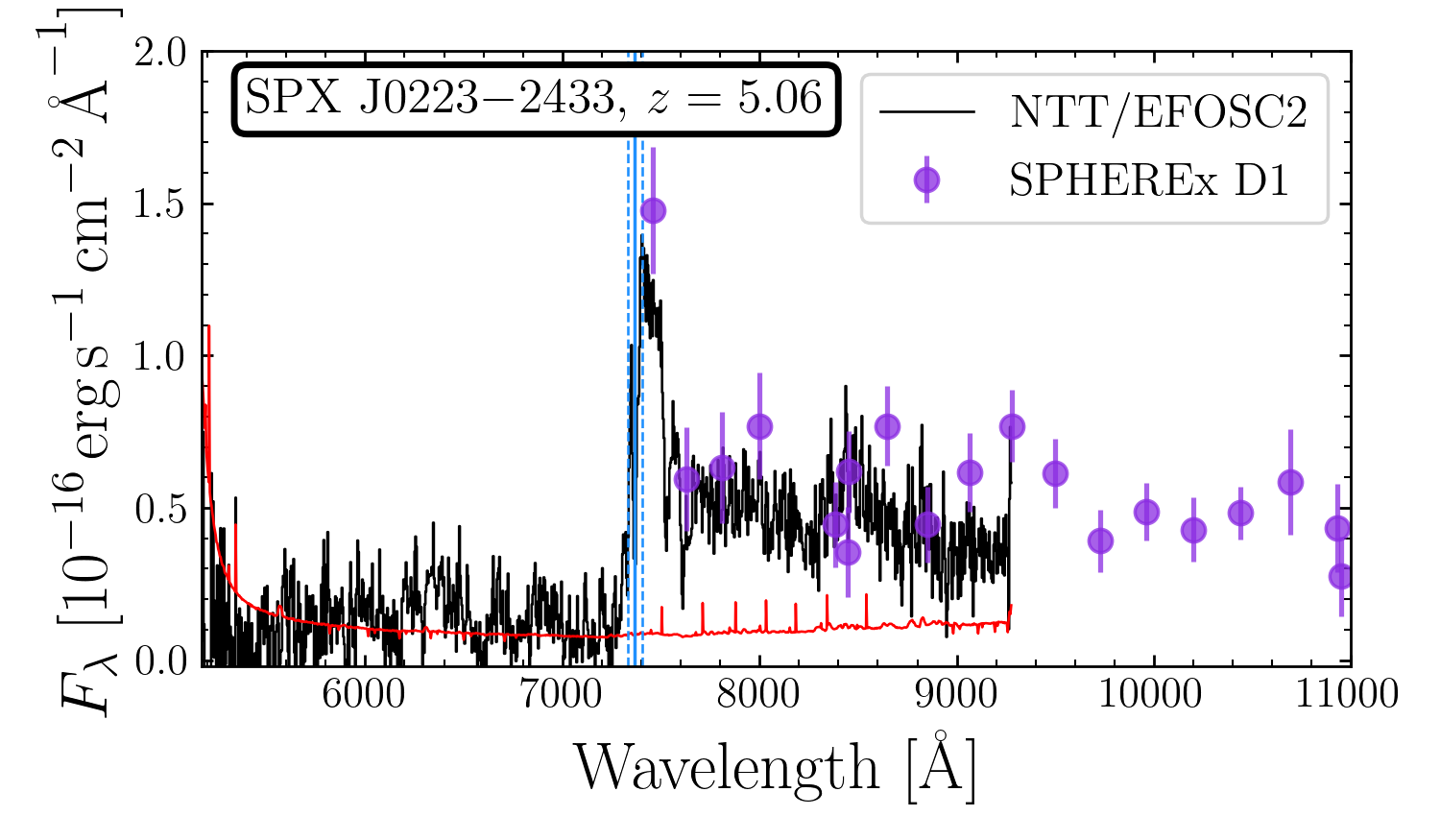}
    \caption{Confirmation spectrum of SPX~J0223$-$2433 taken with NTT/EFOSC2 (black curve) under ESO program 0115.B-2134 (PI Belladitta), approximately two months before our independent identification with SPHEREx.}
    \label{fig:silvia_quasar}
\end{figure}

We estimate the UV absolute magnitude $M_{1450}$ for our new $z>4$ quasars using a template fitting approach. We first fit the \citet{Selsing16} luminous quasar template with attenuation from an SMC extinction law \citep{Gordon03} to the SPHEREx spectrophotometry in a series of line-free spectral regions between rest-frame wavelengths $\lambda_{\rm rest}=1600$\,\AA\ and $\lambda_{\rm rest}=4500$\,\AA. We then evaluate this best-fit model -- including extinction -- at $\lambda_{\rm rest}=1450$\,\AA. Given the relatively poor signal-to-noise ratio of the SPHEREx data at the short wavelength end, we find that this procedure provides a more robust estimate of $M_{1450}$ compare to a direct estimate from the nearest SPHEREx spectrophotometry, and has the added benefit of excluding potential C\,{\small IV} broad absorption features. The estimated $M_{1450}$ values are given in Table~\ref{tab:hiz}.

In Figure~\ref{fig:hiz_known}, we show SPHEREx spectrophotometry for several examples of the known $z>4$ quasars recovered by our visual inspection. Due to the irregular sampling of the SPHEREx survey across the different detectors, for illustrative purposes we re-bin the raw measurements (colored points) onto a fixed wavelength grid (black points) which accounts for the varying spectral sampling between detectors D1--D4 and D5--D6. As the error properties of SPHEREx Quick Release spectrophotometry are not yet fully characterized, these objects highlight what we can expect to see for newly identified quasars. 

\begin{figure}
    \centering
    \includegraphics[width=1.0\linewidth]{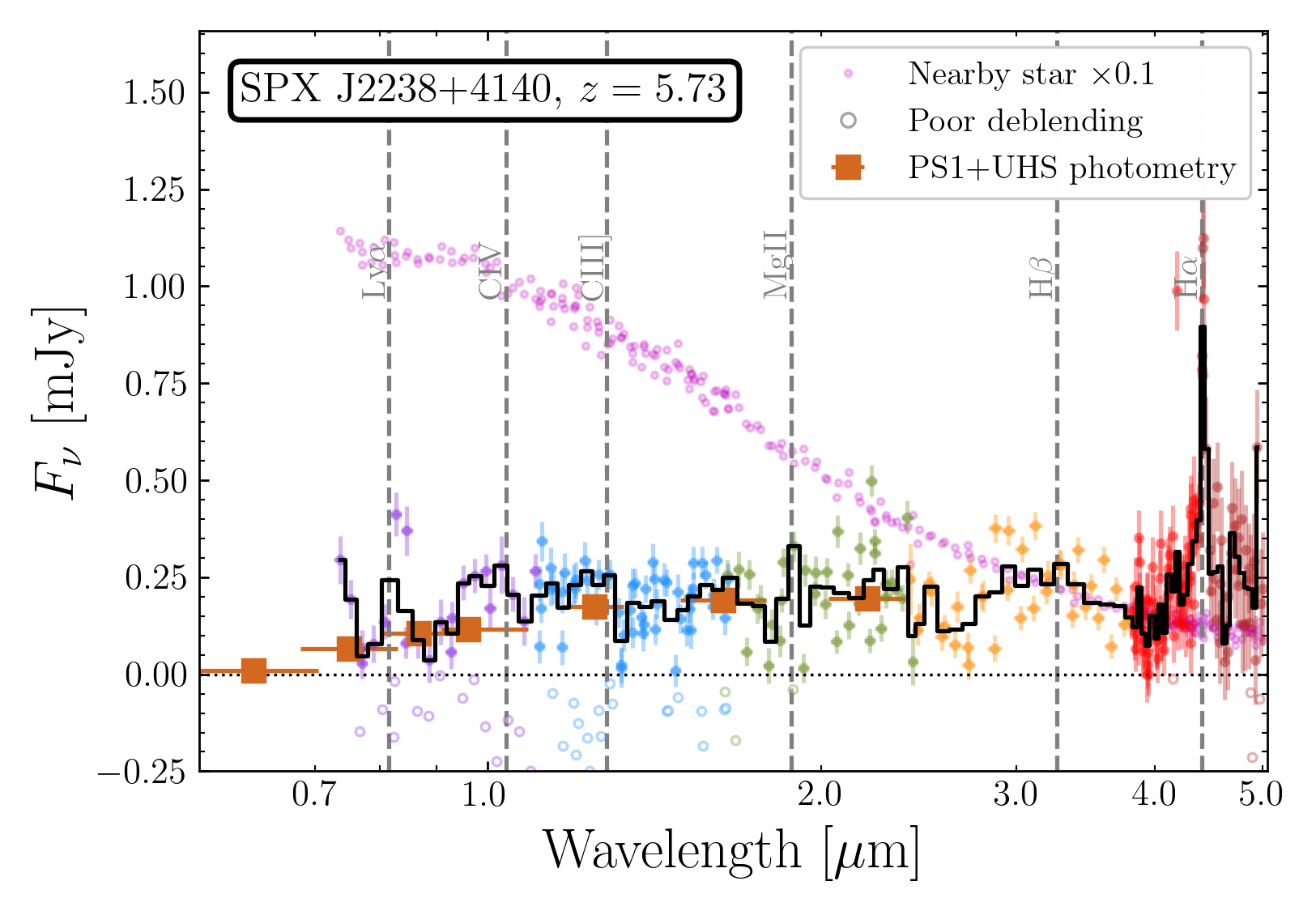}
    \caption{Similar to Figure~\ref{fig:z45_1} but for the newly identified quasar SPX~J2238$+$4140 at $z=5.73$. Negative spectrophotometry values have been masked (open points) from the rebinned spectrum, as they result from poor deblending with a nearby bright object (magenta points, shown divided by 10). Broadband photometry from Pan-STARRS and UHS are shown as brown squares.}
    \label{fig:new_z57}
\end{figure}

We show the SPHEREx spectrophotometry of our new $4<z<5$ quasars in Figure~\ref{fig:z45_1}, and for the new $5<z<5.7$ quasars in Figure~\ref{fig:z5}. Quasars whose coordinates are highlighted in blue are those which have ground-based optical confirmation spectra, either unpublished by other groups as described above or acquired for this work (see Section~\ref{sec:fup}). One example in the former category is SPX~J0223$-$2433, which was independently selected for spectroscopic follow-up prior to the launch of SPHEREx with a method similar to that of \citet{Belladitta25}. In Figure~\ref{fig:silvia_quasar}, we show the ground-based spectrum of J0223$-$2433, taken with the ESO Faint Object Spectrograph and Camera version 2 (EFOSC2; \citealt{Buzzoni84}) on the New Technology Telescope (NTT), which agrees with its SPHEREx-derived redshift. Ten of the remaining unpublished objects were observed with NTT/EFOSC2 by the QUBRICS survey; all of these data have since become publicly available in the ESO archive, and we confirm that these observations all support our SPHEREx identifications. These independent confirmations support the probable high-redshift quasar interpretation of our remaining new quasars.

\begin{table*}[]
    \centering
    \begin{tabular}{lllccl}
    Name & Instrument & Date & Exp. time & Redshift & Notes \\
    \hline \hline
    SPX~J0003$-$1631 & P200/NGPS & 2 Dec 2025 & 1800s & 5.22 &  \\
    SPX~J0021$+$3520 & P200/NGPS & 2 Dec 2025 & 1200s & 4.49 &BAL \\
    SPX~J0200$-$2942 & P200/NGPS & 2 Dec 2025 & 1200s & 5.32&\\
    SPX~J0240$+$3950 & P200/NGPS & 1 Dec 2025 & 1800s & 4.70&  \\
    SPX~J0328$-$1748 & P200/NGPS & 1 Dec 2025 & 1200s & 4.30 &   \\
    SPX~J0413$+$0328 & Keck/KCWI & 29 Dec 2025 & 2160s & 5.26&\\
    SPX~J0533$+$5549 & P200/NGPS & 1 Dec 2025 & 1800s &  4.56& \\
    SPX~J0601$-$2621 & P200/NGPS & 1 Dec 2025 & 1800s & 4.61& H$\alpha$-faintest candidate \\
    SPX~J0701$+$5243 & P200/NGPS & 1 Dec 2025 & 1200s & 5.49 & \\
    SPX~J0709$+$4123 & P200/NGPS & 1 Dec 2025 & 1200s & 4.08 &BAL \\
    SPX~J0717$+$7126 & Keck/KCWI & 29 Dec 2025 & 2400s & 4.83&\\
    SPX~J0930$-$3000 & Keck/KCWI & 29 Dec 2025 & 2400s & 4.58 & \\
    SPX~J1025$-$2443 & P200/NGPS & 1 Dec 2025 & 900s  & 4.94& \\
    SPX~J1152$-$1805 & P200/NGPS & 1 Dec 2025 & 1200s & 4.15 & BAL, unusual/red continuum\\
    SPX~J2040$+$6741 & P200/NGPS & 2 Dec 2025 & 1200s & 4.68&  \\
    SPX~J2109$+$3003 & P200/NGPS & 2 Dec 2025 & 1800s & 4.20& Unusual/red continuum\\
    SPX~J2051$+$2156 & P200/NGPS & 2 Dec 2025 & 1800s & 5.31&BAL  \\
    SPX~J2303$+$3456 & P200/NGPS & 1 Dec 2025 & 1800s & 5.42 & BAL; William's Quasar \\
    \hline
    \end{tabular}
    \caption{Details of the ground-based optical spectroscopic follow-up observations. All SPHEREx-confirmed candidates were found to be $z>4$ quasars, i.e.~the validation success rate was 100\%.}
    \label{tab:followup}
\end{table*}

\begin{figure*}
    \centering
    \includegraphics[width=0.95\linewidth]{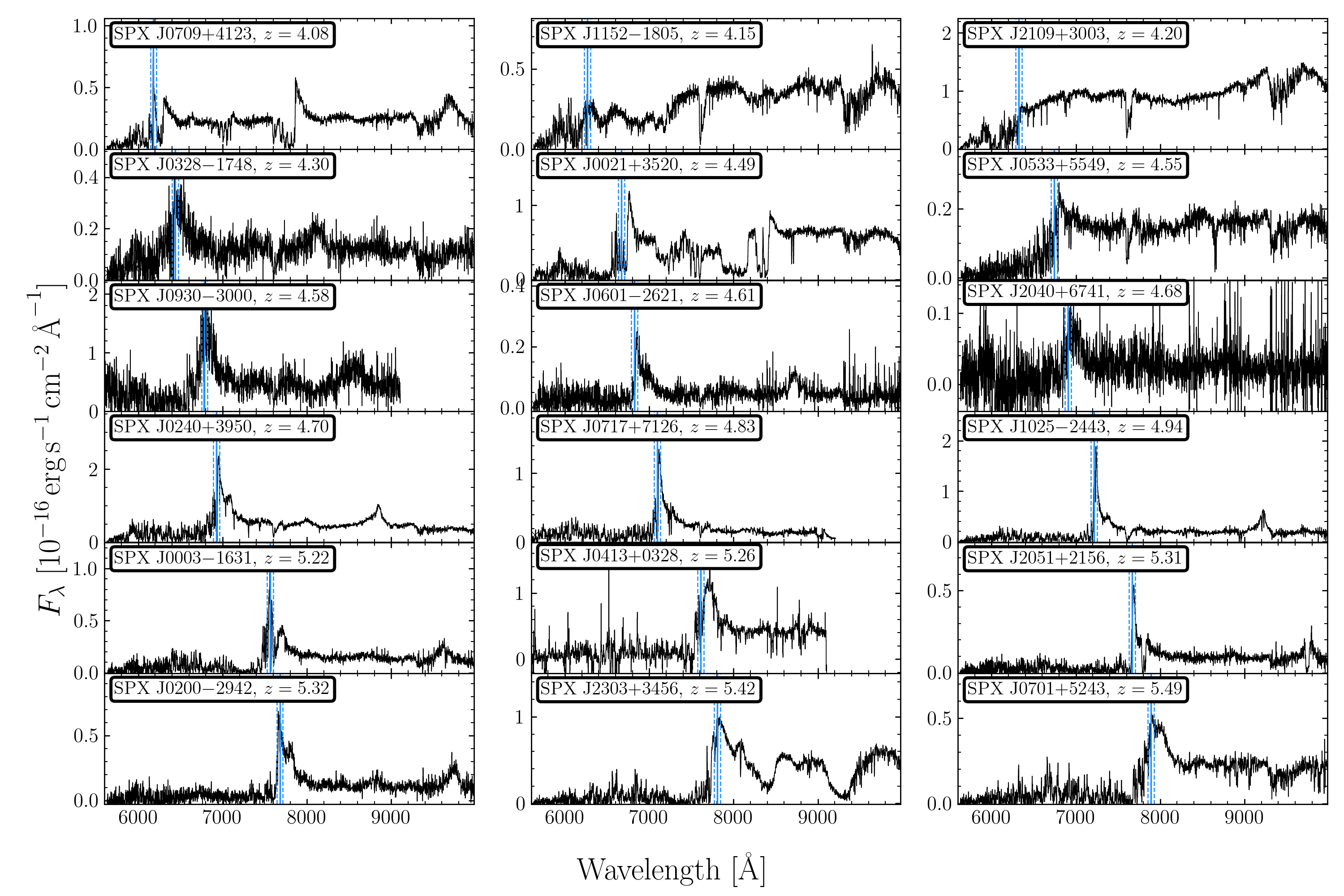}
    \vskip-1em
    \caption{Follow-up spectroscopy of SPHEREx candidates with P200/NGPS and Keck/KCWI. Vertical blue lines show the expected location of Ly$\alpha$ at the nominal SPHEREx H$\alpha$ redshift with our estimated uncertainty of $\delta z=0.03$; the redshifts from ground-based follow-up spectra always agree within this margin.}
    \label{fig:fup}
\end{figure*}

One particularly interesting newly-identified object, SPX~J2238$+$4140, shows a broad emission feature at $\lambda\simeq4.4$\,$\mu$m, corresponding to $z\simeq5.73$ if identified as H$\alpha$. However, most of its spectrophotometry is strongly contaminated by a bright ($F_\nu(1\,\mu{\rm m})\simeq8$\,mJy) object located 8.3 arcsec away. Even after applying the deblending procedure described above, the poor spatial sampling of SPHEREx at its short wavelength end is insufficient to fully separate the two objects, leading to a significant fraction of negative measured fluxes at $\lambda<2$\,$\mu$m. We show the rebinned SPHEREx spectrum in Figure~\ref{fig:new_z57} after masking these negative values. To further validate the spectral shape at the short wavelength end, we also compare to ground-based optical and near-infrared photometry from Pan-STARRS and the UKIDSS Hemisphere Survey (UHS; \citealt{Dye18}), respectively, and find general agreement with the unmasked (i.e.~positive) SPHEREx data. 

\begin{figure*}
    \centering
    \includegraphics[width=0.9\linewidth]{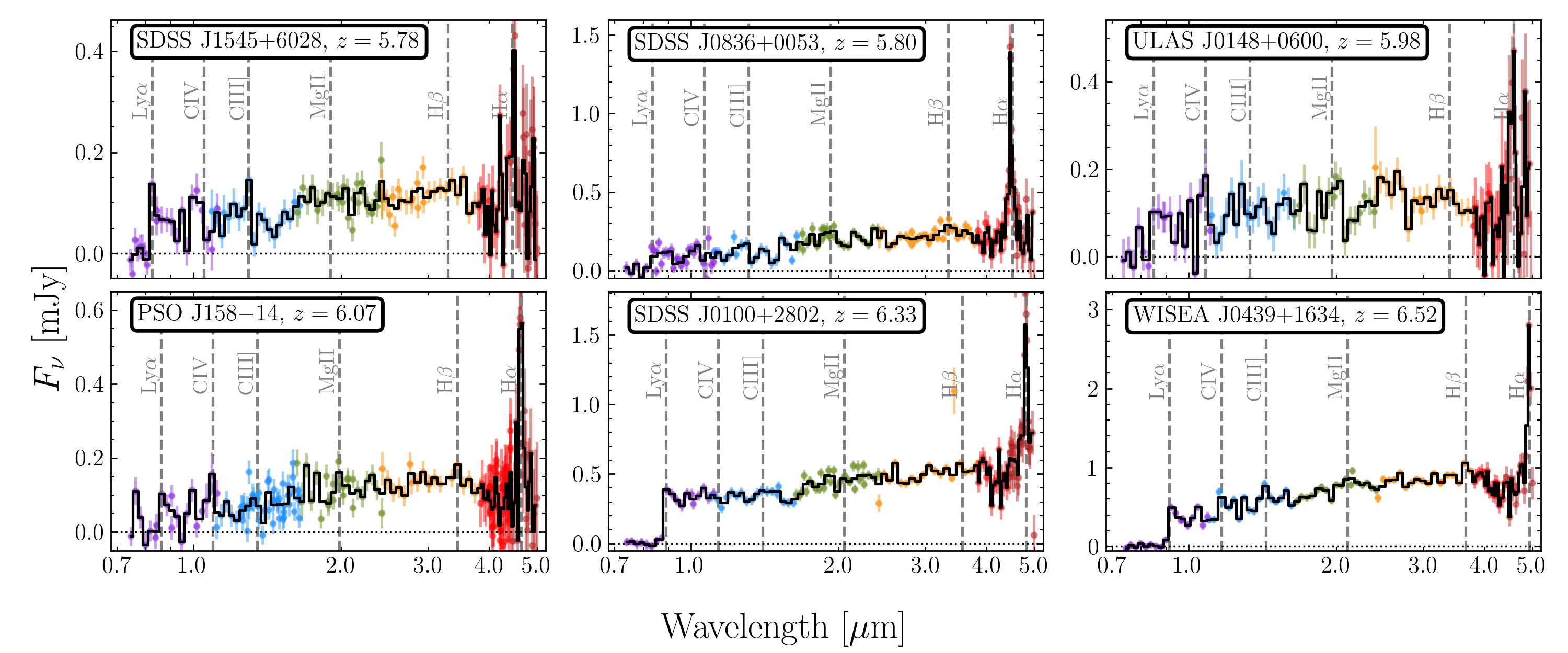}
    \caption{Similar to Figure~\ref{fig:hiz_known}, but now showing example SPHEREx spectrophotometry of previously-known luminous $z>5.7$ quasars.}
    \label{fig:z6spec}
\end{figure*}

\subsection{Ground-based spectroscopic follow-up}\label{sec:fup}

Despite the title and motivation behind this project, and the archival confirmations mentioned previously, the identification of quasars with SPHEREx spectrophotometry is still relatively untested. Thus, for a subset of our new quasars, we obtained our own ground-based follow-up spectroscopy for validation. Because our sample is relatively bright, their confirmation is an ideal backup program for relatively poor weather conditions on large telescopes. 

On 1 Dec 2025, 2 Dec 2025, and 19 Dec 2025 we observed 16 of our new $z>4$ quasars with the Next Generation Palomar Spectrograph (NGPS; \citealt{NGPS}) on the 200-inch Hale telescope at Palomar Observatory, in bright sky conditions, intermittent clouds, and 1-1.5 arcsec seeing. 

On 29 Dec 2025, two more $z>4$ quasars were observed with the Keck Cosmic Web Imager (KCWI; \citealt{KCWI}) on the Keck II telescope. Observations were conducted using the medium slicer configuration, which provides a field of view of 16\farcs5\,$\times$\,20\farcs4, composed of 24 slices each with a slice width of 0\farcs69. The RL grating was employed in the red arm, covering 5550\,--9239\,\AA. A binning of $2 \times 2$ was applied. The spectral resolution of this setup is ${\rm FWHM} \approx 300\,{\rm km/s}$ ($R\sim 1000$). 

Additional observational details are provided in Table~\ref{tab:followup}. As shown in Figure~\ref{fig:fup}, all 18 quasar candidates were confirmed, with redshifts consistent with our SPHEREx determinations upon visual inspection. This extremely high confirmation rate, even for one of our faintest targets with $G_\text{RP}=19.40$, SPX~J0601$-$2621, suggests that identification with SPHEREx alone is likely sufficient to confirm quasars and their spectroscopic redshifts.


\subsection{Lower redshift quasars}
Due to the lack of an upper limit to $W1-W2$ color, our fiducial selection does not exclude lower redshift quasars, where a particularly red color can result from the onset of hot ($T\gtrsim1000$\,K) dust emission from the torus at rest-frame wavelengths redward of 1\,$\mu$m (e.g. \citealt{Hernan16, Bosman25}). As a result, we identify \lowztot previously unknown lower redshift quasars at $0.4 \lesssim z \lesssim 4$ in the fiducial selection. In contrast to the higher redshift sample, we only recover \lowzknown known quasars in the same redshift interval. The majority of the newly identified quasars lie at $1.2 \leq z \leq 2.0$, and likely correspond to intrinsically or Galactically reddened and/or FeLoBAL quasars, where the unusually red \emph{Gaia} color results from dust reddening or strong Mg\,{\small II} and/or Fe\,{\small II} absorption, respectively.

In Figure~\ref{fig:lowz} we show six examples of newly identified $z<4$ quasars, including a luminous ($M_{3000}\simeq-29$) quasar at $z=3.46$, SPX~J0048+8328. This quasar exhibits a clear break below rest-frame Mg\,{\small II} $\lambda$2800, likely due to strong FeLoBAL absorption from Mg\,{\small II} as well as the dense forest of Fe\,{\small II} transitions.

\section{A bright future for SPHEREx at high-redshift}
\label{sec:future}

Our quasar selection is generally limited to discoveries at $z\lesssim5.5$ due to technical limitations of \emph{Gaia} DR3 processing. However, even higher redshift quasars are in principle detectable with SPHEREx: redshifted H$\alpha$ is accessible up to $z=6.6$, and the Ly$\alpha$ break is theoretically available for all $z\gtrsim5.2$ objects (e.g. Figure~\ref{fig:j0306}). In Figure~\ref{fig:z6spec}, we show SPHEREx spectrophotometry of a selection of some of the brightest known $z>5.7$ quasars. The first four panels demonstrate that quasars of $\sim19\!-\!20$ AB apparent magnitude at $z\sim6$ already show detections of H$\alpha$ and (to a lesser extent) Ly$\alpha$ breaks in the first SPHEREx all-sky survey. The bottom two panels show the two brightest objects known at $z>6$: the luminous quasar SDSS~J0100+2802 \citep{Wu15} and the strongly-lensed quasar J0439+1634 \citep{Fan19}. For these objects, not only are their H$\alpha$ and other broad lines prominently visible, their Ly$\alpha$ breaks are clearly detected. 

The final SPHEREx data release will provide similar-quality data as Figure~\ref{fig:z6spec} for objects a factor of two fainter, enabling a nearly complete all-sky census of the brightest (most luminous and/or strongly lensed) objects at arbitrarily high redshift. In addition, the SPHEREx deep fields close to the northern and southern ecliptic poles will have $\sim100\times$ as many visits, implying similar quality detections of continuum breaks a few magnitudes fainter should be possible, albeit over only a few hundred square degrees of the sky.

\section{Conclusion}
\label{sec:ccl}

In this work, we have investigated the potential for SPHEREx to discover luminous high-redshift quasars. 
We identified \hiznew new luminous quasars at $z > 4$, including \hiznewb at $z > 5$, primarily from the detection of broad H$\alpha$ emission lines in SPHEREx spectrophotometry of \emph{Gaia}+\emph{WISE}-selected candidates. Ground-based spectroscopic follow-up has confirmed 18/18 targets (i.e. a confirmation rate of 100\%), suggesting that our visual inspection of SPHEREx data alone is sufficient to confirm luminous high-redshift quasars with high purity. In addition, we identify \lowznew new quasars at $z < 4$, likely consisting predominantly of reddened and BAL quasars with unusual optical colors.

For the vast majority of the previously known quasars we recover, this is the first time that their rest-optical properties have been probed beyond the existing broadband WISE photometry. For those quasars whose H$\alpha$ emission lines lie redward of 3.8\,$\mu$m, i.e. at $z\gtrsim4.8$, the effective spectral resolution of the SPHEREx spectrophotometry may be sufficient to measure the intrinsic width of the line, enabling measurement of their black hole masses via a single-epoch virial estimator \citep{DallaBonta25}. Such measurements will benefit greatly from the \emph{third} all-sky survey, scheduled to be completed at the end of 2026, which will deliver Nyquist sampled spectrophotometry for all sources. For quasars with particularly broad H$\alpha$ emission lines, the spectrophotometry used in this work may already be up to this task, but we defer such analysis to future work, as it will require a careful accounting of the effective line spread function of the SPHEREx LVF.

While we have relied on photometry and astrometry from previous surveys to select quasar candidates, the SPHEREx mission will eventually deliver all-sky spectrophotometry that will allow for comprehensive emission line and continuum break searches irrespective of ancillary data. Such an undertaking will surely result in a more complete and unbiased sample of high-redshift quasars across the full sky, enabling statistical studies of the quasar luminosity function and environmental dependence that are currently limited by the heterogeneous selection criteria of targeted surveys. 

\begin{acknowledgements}

    SEIB and AG are supported by the Deutsche Forschungsgemeinschaft (DFG) under Emmy Noether grant number BO 5771/1-1.

    This publication makes use of data products from the Spectro-Photometer for the History of the Universe, Epoch of Reionization and Ices Explorer (SPHEREx), which is a joint project of the Jet Propulsion Laboratory and the California Institute of Technology, and is funded by the National Aeronautics and Space Administration. 

    This work has made use of data from the European Space Agency (ESA) mission
{\it Gaia} (\url{https://www.cosmos.esa.int/gaia}), processed by the {\it Gaia}
Data Processing and Analysis Consortium (DPAC,
\url{https://www.cosmos.esa.int/web/gaia/dpac/consortium}). Funding for the DPAC
has been provided by national institutions, in particular the institutions
participating in the {\it Gaia} Multilateral Agreement.

    This publication makes use of data products from the Wide-field Infrared Survey Explorer, which is a joint project of the University of California, Los Angeles, and the Jet Propulsion Laboratory/California Institute of Technology, funded by the National Aeronautics and Space Administration.

    This publication also makes use of data products from NEOWISE, which is a project of the Jet Propulsion Laboratory/California Institute of Technology, funded by the Planetary Science Division of the National Aeronautics and Space Administration.

    This research has made use of the NASA/IPAC Infrared Science Archive, which is funded by the National Aeronautics and Space Administration and operated by the California Institute of Technology.

    This research has made use of ``Aladin sky atlas'' developed at CDS, Strasbourg Observatory, France. 

    Based on data obtained from the ESO Science Archive Facility.

    Based on observations obtained at the Hale Telescope, Palomar Observatory as part of a continuing collaboration between the California Institute of Technology, NASA/JPL, Yale University, and the National Astronomical Observatories of China.

    Some of the data presented herein were obtained at Keck Observatory, which is a private 501(c)3 non-profit organization operated as a scientific partnership among the California Institute of Technology, the University of California, and the National Aeronautics and Space Administration. The Observatory was made possible by the generous financial support of the W. M. Keck Foundation. 

    The authors wish to recognize and acknowledge the very significant cultural role and reverence that the summit of Maunakea has always had within the Native Hawaiian community. We are most fortunate to have the opportunity to conduct observations from this mountain.

\end{acknowledgements}

\bibliographystyle{aa} 

\begin{thebibliography}{67}
\expandafter\ifx\csname natexlab\endcsname\relax\def\natexlab#1{#1}\fi

\bibitem[{{Ba{\~n}ados} {et~al.}(2025){Ba{\~n}ados}, {Le Brun}, {Belladitta},
  {Momcheva}, {Stern}, {Wolf}, {Ezziati}, {Mortlock}, {Humphrey}, {Smart},
  {Casewell}, {P{\'e}rez-Garrido}, {Goldman}, {Mart{\'\i}n}, {Mohandasan},
  {Reyl{\'e}}, {Dominguez-Tagle}, {Copin}, {Lusso}, {Matsuoka}, {McCarthy},
  {Ricci}, {Rix}, {Rottgering}, {Schindler}, {Weaver}, {Allaoui}, {Bedrine},
  {Castellano}, {Chabaud}, {Daste}, {Dufresne}, {Gracia-Carpio}, {K{\"u}mmel},
  {Moresco}, {Scodeggio}, {Surace}, {Vibert}, {Balestra}, {Bonnefoi},
  {Caillat}, {Cogato}, {Costille}, {Dusini}, {Ferriol}, {Franceschi},
  {Gillard}, {Jahnke}, {Le Mignant}, {Ligori}, {Medinaceli}, {Morgante},
  {Passalacqua}, {Paterson}, {Pires}, {Sirignano}, {Andika}, {Atek}, {Barrado},
  {Bisogni}, {Conselice}, {Dannerbauer}, {Decarli}, {Dole}, {Dupuy}, {Feltre},
  {Fotopoulou}, {Gillis}, {Lopez}, {Onoue}, {Rodighiero}, {Sedighi}, {Shankar},
  {Siudek}, {Spinoglio}, {Vergani}, {Vietri}, {Walter}, {Zamorani}, {Zapatero
  Osorio}, {Zhang}, {Bethermin}, {Aghanim}, {Altieri}, {Amara}, {Andreon},
  {Baccigalupi}, {Baldi}, {Bardelli}, {Basset}, {Battaglia}, {Biviano},
  {Bonchi}, {Bonino}, {Branchini}, {Brescia}, {Brinchmann}, {Camera},
  {Capobianco}, {Carbone}, {Carretero}, {Casas}, {Castignani}, {Cavuoti},
  {Cimatti}, {Colodro-Conde}, {Congedo}, {Conversi}, {Courbin}, {Courtois},
  {Cropper}, {Cuby}, {Da Silva}, {Degaudenzi}, {De Lucia}, {Giorgio},
  {Dolding}, {Dubath}, {Duncan}, {Dupac}, {Ealet}, {Farina}, {Faustini},
  {Fourmanoit}, {Frailis}, {Galeotta}, {George}, {Giocoli}, {Granett},
  {Grazian}, {Grupp}, {Guzzo}, {Haugan}, {Hoar}, {Hoekstra}, {Holmes}, {Hook},
  {Hormuth}, {Hornstrup}, {Hudelot}, {Jhabvala}, {Joachimi}, {Keih{\"a}nen},
  {Kermiche}, {Kubik}, {Kuijken}, {Kunz}, {Kurki-Suonio}, {Lilje}, {Lindholm},
  {Lloro}, {Mainetti}, {Maino}, {Maiorano}, {Mansutti}, {Marggraf}, {Markovic},
  {Martinelli}, {Martinet}, {Marulli}, {Massey}, {Mei}, {Mellier},
  {Meneghetti}, {Merlin}, {Meylan}, {Mora}, {Moscardini}, {Neissner}, {Niemi},
  {Nightingale}, {Padilla}, {Paltani}, {Pasian}, {Pedersen}, {Percival},
  {Pettorino}, {Polenta}, {Poncet}, {Popa}, {Pozzetti}, {Raison}, {Rebolo},
  {Renzi}, {Rhodes}, {Riccio}, {Romelli}, {Roncarelli}, {Rossetti}, {Saglia},
  {Sakr}, {Sapone}, {Sartoris}, {Schewtschenko}, {Schirmer}, {Schneider},
  {Schrabback}, {Secroun}, {Sefusatti}, \& {Seidel}}]{Banados25}
{Ba{\~n}ados}, E., {Le Brun}, V., {Belladitta}, S., {et~al.} 2025, \mnras, 542,
  1088

\bibitem[{{Ba{\~n}ados} {et~al.}(2023){Ba{\~n}ados}, {Schindler}, {Venemans},
  {Connor}, {Decarli}, {Farina}, {Mazzucchelli}, {Meyer}, {Stern}, {Walter},
  {Fan}, {Hennawi}, {Khusanova}, {Morrell}, {Nanni}, {Noirot}, {Pensabene},
  {Rix}, {Simon}, {Verdoes Kleijn}, {Xie}, {Yang}, \& {Connor}}]{Banados23}
{Ba{\~n}ados}, E., {Schindler}, J.-T., {Venemans}, B.~P., {et~al.} 2023, \apjs,
  265, 29

\bibitem[{{Ba{\~n}ados} {et~al.}(2016){Ba{\~n}ados}, {Venemans}, {Decarli},
  {Farina}, {Mazzucchelli}, {Walter}, {Fan}, {Stern}, {Schlafly}, {Chambers},
  {Rix}, {Jiang}, {McGreer}, {Simcoe}, {Wang}, {Yang}, {Morganson}, {De Rosa},
  {Greiner}, {Balokovi{\'c}}, {Burgett}, {Cooper}, {Draper}, {Flewelling},
  {Hodapp}, {Jun}, {Kaiser}, {Kudritzki}, {Magnier}, {Metcalfe}, {Miller},
  {Schindler}, {Tonry}, {Wainscoat}, {Waters}, \& {Yang}}]{Banados16}
{Ba{\~n}ados}, E., {Venemans}, B.~P., {Decarli}, R., {et~al.} 2016, \apjs, 227,
  11

\bibitem[{{Becker} {et~al.}(2015){Becker}, {Bolton}, {Madau}, {Pettini},
  {Ryan-Weber}, \& {Venemans}}]{Becker15}
{Becker}, G.~D., {Bolton}, J.~S., {Madau}, P., {et~al.} 2015, \mnras, 447, 3402

\bibitem[{{Becker} {et~al.}(2021){Becker}, {D'Aloisio}, {Christenson}, {Zhu},
  {Worseck}, \& {Bolton}}]{Becker21}
{Becker}, G.~D., {D'Aloisio}, A., {Christenson}, H.~M., {et~al.} 2021, \mnras,
  508, 1853

\bibitem[{{Belladitta} {et~al.}(2025){Belladitta}, {Ba{\~n}ados}, {Xie},
  {Decarli}, {Onorato}, {Yang}, {Bischetti}, {Onoue}, {Loiacono},
  {Mart{\'\i}nez-Ram{\'\i}rez}, {Mazzucchelli}, {Davies}, {Wolf}, {Schindler},
  {Fan}, {Wang}, {Walter}, {Mkrtchyan}, {Stern}, {Farina}, \&
  {Venemans}}]{Belladitta25}
{Belladitta}, S., {Ba{\~n}ados}, E., {Xie}, Z.-L., {et~al.} 2025, \aap, 699,
  A335

\bibitem[{{Bock} {et~al.}(2025){Bock}, {Aboobaker}, {Adamo}, {Akeson}, {Alred},
  {Alibay}, {Ashby}, {Bach}, {Bleem}, {Bolton}, {Braun}, {Bruton}, {Bryan},
  {Chang}, {Chen}, {Cheng}, {Cheshire}, {Chiang}, {Choppin de Janvry},
  {Condon}, {Cook}, {Cooray}, {Crill}, {Cukierman}, {Dore}, {Dowell},
  {Dubois-Felsmann}, {Eifler}, {Everett}, {Fabinsky}, {Faisst}, {Fanson},
  {Farrington}, {Fatahi}, {Fazar}, {Feder}, {Frater}, {Grasshorn Gebhardt},
  {Giri}, {Goldina}, {Gorjian}, {Habib}, {Hart}, {Heinrich}, {Hora}, {Huai},
  {Hui}, {Jo}, {Jeong}, {Kang}, {Kang}, {Kecman}, {Kim}, {Kim}, {Kim}, {Kim},
  {Kim}, {Kirkpatrick}, {Kobayashi}, {Korngut}, {Krause}, {Lee}, {Lee}, {Lee},
  {Lee}, {Lisse}, {Mariani}, {Masters}, {Mauskopf}, {Melnick}, {Minasyan},
  {Mirocha}, {Miyasaka}, {Moore}, {Moore}, {Murgia}, {Naylor}, {Nelson},
  {Nguyen}, {Nguyen}, {Noh}, {Padin}, {Paladini}, {Park}, {Penanen}, {Putnam},
  {Pyo}, {Ramachandra}, {Ramanathan}, {Rustamkulov}, {Reiley}, {Rice}, {Rocca},
  {Seok}, {Smith}, {Stober}, {Susca}, {Teplitz}, {Thelen}, {Tolls}, {Torrini},
  {Trangsrud}, {Unwin}, {Velicheti}, {Wang}, {Wen}, {-Werner}, {Williams},
  {Williamson}, {Wincentsen}, {Windhorst}, {Yang}, {Yang}, \&
  {Zemcov}}]{Bock25}
{Bock}, J.~J., {Aboobaker}, A.~M., {Adamo}, J., {et~al.} 2025, arXiv e-prints,
  arXiv:2511.02985

\bibitem[{{Boera} {et~al.}(2019){Boera}, {Becker}, {Bolton}, \&
  {Nasir}}]{Boera19}
{Boera}, E., {Becker}, G.~D., {Bolton}, J.~S., \& {Nasir}, F. 2019, \apj, 872,
  101

\bibitem[{{Bonnarel} {et~al.}(2000){Bonnarel}, {Fernique}, {Bienaym{\'e}},
  {Egret}, {Genova}, {Louys}, {Ochsenbein}, {Wenger}, \&
  {Bartlett}}]{Bonnarel00}
{Bonnarel}, F., {Fernique}, P., {Bienaym{\'e}}, O., {et~al.} 2000, \aaps, 143,
  33

\bibitem[{{Bosman} {et~al.}(2025){Bosman}, {{\'A}lvarez-M{\'a}rquez}, {Davies},
  {Protu{\v{s}}ov{\'a}}, {Hennawi}, {Yang}, {Spina}, {Colina}, {Fan},
  {{\"O}stlin}, {Walter}, {Wang}, {Ward}, {Alonso Herrero}, {Barth},
  {Belladitta}, {Boogaard}, {Caputi}, {Connor},
  {{\v{D}}urov{\v{c}}{\'\i}kov{\'a}}, {Eilers}, {Crespo G{\'o}mez}, {Hjorth},
  {Jun}, {Langeroodi}, {Liu}, {Lupi}, {Mazzucchelli}, {Pye}, {Rinaldi}, {van
  der Werf}, \& {Volonteri}}]{Bosman25}
{Bosman}, S. E.~I., {{\'A}lvarez-M{\'a}rquez}, J., {Davies}, F.~B., {et~al.}
  2025, arXiv e-prints, arXiv:2511.02902

\bibitem[{{Bosman} {et~al.}(2022){Bosman}, {Davies}, {Becker}, {Keating},
  {Davies}, {Zhu}, {Eilers}, {D'Odorico}, {Bian}, {Bischetti}, {Cristiani},
  {Fan}, {Farina}, {Haehnelt}, {Hennawi}, {Kulkarni}, {Mesinger}, {Meyer},
  {Onoue}, {Pallottini}, {Qin}, {Ryan-Weber}, {Schindler}, {Walter}, {Wang}, \&
  {Yang}}]{Bosman22}
{Bosman}, S. E.~I., {Davies}, F.~B., {Becker}, G.~D., {et~al.} 2022, \mnras,
  514, 55

\bibitem[{{Brazzini} {et~al.}(2025){Brazzini}, {D'Odorico}, {Bischetti},
  {Feruglio}, {Cupani}, {Becker}, \& {Tripodi}}]{Brazzini25}
{Brazzini}, M., {D'Odorico}, V., {Bischetti}, M., {et~al.} 2025, \aap, 698,
  A145

\bibitem[{{Buzzoni} {et~al.}(1984){Buzzoni}, {Delabre}, {Dekker}, {Dodorico},
  {Enard}, {Focardi}, {Gustafsson}, {Nees}, {Paureau}, \& {Reiss}}]{Buzzoni84}
{Buzzoni}, B., {Delabre}, B., {Dekker}, H., {et~al.} 1984, The Messenger, 38, 9

\bibitem[{{Chambers} {et~al.}(2016){Chambers}, {Magnier}, {Metcalfe},
  {Flewelling}, {Huber}, {Waters}, {Denneau}, {Draper}, {Farrow}, {Finkbeiner},
  {Holmberg}, {Koppenhoefer}, {Price}, {Rest}, {Saglia}, {Schlafly}, {Smartt},
  {Sweeney}, {Wainscoat}, {Burgett}, {Chastel}, {Grav}, {Heasley}, {Hodapp},
  {Jedicke}, {Kaiser}, {Kudritzki}, {Luppino}, {Lupton}, {Monet}, {Morgan},
  {Onaka}, {Shiao}, {Stubbs}, {Tonry}, {White}, {Ba{\~n}ados}, {Bell},
  {Bender}, {Bernard}, {Boegner}, {Boffi}, {Botticella}, {Calamida},
  {Casertano}, {Chen}, {Chen}, {Cole}, {Deacon}, {Frenk}, {Fitzsimmons},
  {Gezari}, {Gibbs}, {Goessl}, {Goggia}, {Gourgue}, {Goldman}, {Grant},
  {Grebel}, {Hambly}, {Hasinger}, {Heavens}, {Heckman}, {Henderson}, {Henning},
  {Holman}, {Hopp}, {Ip}, {Isani}, {Jackson}, {Keyes}, {Koekemoer}, {Kotak},
  {Le}, {Liska}, {Long}, {Lucey}, {Liu}, {Martin}, {Masci}, {McLean}, {Mindel},
  {Misra}, {Morganson}, {Murphy}, {Obaika}, {Narayan}, {Nieto-Santisteban},
  {Norberg}, {Peacock}, {Pier}, {Postman}, {Primak}, {Rae}, {Rai}, {Riess},
  {Riffeser}, {Rix}, {R{\"o}ser}, {Russel}, {Rutz}, {Schilbach}, {Schultz},
  {Scolnic}, {Strolger}, {Szalay}, {Seitz}, {Small}, {Smith}, {Soderblom},
  {Taylor}, {Thomson}, {Taylor}, {Thakar}, {Thiel}, {Thilker}, {Unger},
  {Urata}, {Valenti}, {Wagner}, {Walder}, {Walter}, {Watters}, {Werner},
  {Wood-Vasey}, \& {Wyse}}]{Chambers16}
{Chambers}, K.~C., {Magnier}, E.~A., {Metcalfe}, N., {et~al.} 2016, arXiv
  e-prints, arXiv:1612.05560

\bibitem[{{Chaussidon} {et~al.}(2023){Chaussidon}, {Y{\`e}che},
  {Palanque-Delabrouille}, {Alexander}, {Yang}, {Ahlen}, {Bailey}, {Brooks},
  {Cai}, {Chabanier}, {Davis}, {Dawson}, {de laMacorra}, {Dey}, {Dey},
  {Eftekharzadeh}, {Eisenstein}, {Fanning}, {Font-Ribera}, {Gazta{\~n}aga}, {A
  Gontcho}, {Gonzalez-Morales}, {Guy}, {Herrera-Alcantar}, {Honscheid},
  {Ishak}, {Jiang}, {Juneau}, {Kehoe}, {Kisner}, {Kov{\'a}cs}, {Kremin}, {Lan},
  {Landriau}, {Le Guillou}, {Levi}, {Magneville}, {Martini}, {Meisner},
  {Moustakas}, {Mu{\~n}oz-Guti{\'e}rrez}, {Myers}, {Newman}, {Nie}, {Percival},
  {Poppett}, {Prada}, {Raichoor}, {Ravoux}, {Ross}, {Schlafly}, {Schlegel},
  {Tan}, {Tarl{\'e}}, {Zhou}, {Zhou}, \& {Zou}}]{DESI_qso}
{Chaussidon}, E., {Y{\`e}che}, C., {Palanque-Delabrouille}, N., {et~al.} 2023,
  \apj, 944, 107

\bibitem[{{Cristiani} {et~al.}(2023){Cristiani}, {Porru}, {Guarneri},
  {Calderone}, {Boutsia}, {Grazian}, {Cupani}, {D'Odorico}, {Fontanot},
  {Martins}, {Marques}, {Maitra}, \& {Trost}}]{Cristiani23}
{Cristiani}, S., {Porru}, M., {Guarneri}, F., {et~al.} 2023, \mnras, 522, 2019

\bibitem[{{Dalla Bont{\`a}} {et~al.}(2025){Dalla Bont{\`a}}, {Peterson},
  {Grier}, {Berton}, {Brandt}, {Ciroi}, {Corsini}, {Dalla Barba}, {Davies},
  {Dehghanian}, {Edelson}, {Foschini}, {Gasparri}, {Ho}, {Horne}, {Iodice},
  {Morelli}, {Pizzella}, {Portaluri}, {Shen}, {Schneider}, \&
  {Vestergaard}}]{DallaBonta25}
{Dalla Bont{\`a}}, E., {Peterson}, B.~M., {Grier}, C.~J., {et~al.} 2025, \aap,
  696, A48

\bibitem[{{Davies} {et~al.}(2018){Davies}, {Hennawi}, {Ba{\~n}ados},
  {Luki{\'c}}, {Decarli}, {Fan}, {Farina}, {Mazzucchelli}, {Rix}, {Venemans},
  {Walter}, {Wang}, \& {Yang}}]{Davies18b}
{Davies}, F.~B., {Hennawi}, J.~F., {Ba{\~n}ados}, E., {et~al.} 2018, \apj, 864,
  142

\bibitem[{{Davies} {et~al.}(2023){Davies}, {Ryan-Weber}, {D'Odorico}, {Bosman},
  {Meyer}, {Becker}, {Cupani}, {Keating}, {Bischetti}, {Davies}, {Eilers},
  {Farina}, {Haehnelt}, {Pallottini}, \& {Zhu}}]{RDavies23}
{Davies}, R.~L., {Ryan-Weber}, E., {D'Odorico}, V., {et~al.} 2023, \mnras, 521,
  314

\bibitem[{{Dey} {et~al.}(2019){Dey}, {Schlegel}, {Lang}, {Blum}, {Burleigh},
  {Fan}, {Findlay}, {Finkbeiner}, {Herrera}, {Juneau}, {Landriau}, {Levi},
  {McGreer}, {Meisner}, {Myers}, {Moustakas}, {Nugent}, {Patej}, {Schlafly},
  {Walker}, {Valdes}, {Weaver}, {Y{\`e}che}, {Zou}, {Zhou}, {Abareshi},
  {Abbott}, {Abolfathi}, {Aguilera}, {Alam}, {Allen}, {Alvarez}, {Annis},
  {Ansarinejad}, {Aubert}, {Beechert}, {Bell}, {BenZvi}, {Beutler}, {Bielby},
  {Bolton}, {Brice{\~n}o}, {Buckley-Geer}, {Butler}, {Calamida}, {Carlberg},
  {Carter}, {Casas}, {Castander}, {Choi}, {Comparat}, {Cukanovaite}, {Delubac},
  {DeVries}, {Dey}, {Dhungana}, {Dickinson}, {Ding}, {Donaldson}, {Duan},
  {Duckworth}, {Eftekharzadeh}, {Eisenstein}, {Etourneau}, {Fagrelius},
  {Farihi}, {Fitzpatrick}, {Font-Ribera}, {Fulmer}, {G{\"a}nsicke},
  {Gaztanaga}, {George}, {Gerdes}, {Gontcho}, {Gorgoni}, {Green}, {Guy},
  {Harmer}, {Hernandez}, {Honscheid}, {Huang}, {James}, {Jannuzi}, {Jiang},
  {Joyce}, {Karcher}, {Karkar}, {Kehoe}, {Kneib}, {Kueter-Young}, {Lan},
  {Lauer}, {Le Guillou}, {Le Van Suu}, {Lee}, {Lesser}, {Perreault Levasseur},
  {Li}, {Mann}, {Marshall}, {Mart{\'\i}nez-V{\'a}zquez}, {Martini}, {du Mas des
  Bourboux}, {McManus}, {Meier}, {M{\'e}nard}, {Metcalfe},
  {Mu{\~n}oz-Guti{\'e}rrez}, {Najita}, {Napier}, {Narayan}, {Newman}, {Nie},
  {Nord}, {Norman}, {Olsen}, {Paat}, {Palanque-Delabrouille}, {Peng},
  {Poppett}, {Poremba}, {Prakash}, {Rabinowitz}, {Raichoor}, {Rezaie},
  {Robertson}, {Roe}, {Ross}, {Ross}, {Rudnick}, {Safonova}, {Saha},
  {S{\'a}nchez}, {Savary}, {Schweiker}, {Scott}, {Seo}, {Shan}, {Silva},
  {Slepian}, {Soto}, {Sprayberry}, {Staten}, {Stillman}, {Stupak}, {Summers},
  {Sien Tie}, {Tirado}, {Vargas-Maga{\~n}a}, {Vivas}, {Wechsler}, {Williams},
  {Yang}, {Yang}, {Yapici}, {Zaritsky}, {Zenteno}, {Zhang}, {Zhang}, {Zhou}, \&
  {Zhou}}]{Dey19}
{Dey}, A., {Schlegel}, D.~J., {Lang}, D., {et~al.} 2019, \aj, 157, 168

\bibitem[{{Dor{\'e}} {et~al.}(2014){Dor{\'e}}, {Bock}, {Ashby}, {Capak},
  {Cooray}, {de Putter}, {Eifler}, {Flagey}, {Gong}, {Habib}, {Heitmann},
  {Hirata}, {Jeong}, {Katti}, {Korngut}, {Krause}, {Lee}, {Masters},
  {Mauskopf}, {Melnick}, {Mennesson}, {Nguyen}, {{\"O}berg}, {Pullen},
  {Raccanelli}, {Smith}, {Song}, {Tolls}, {Unwin}, {Venumadhav}, {Viero},
  {Werner}, \& {Zemcov}}]{SPHEREx}
{Dor{\'e}}, O., {Bock}, J., {Ashby}, M., {et~al.} 2014, arXiv e-prints,
  arXiv:1412.4872

\bibitem[{{Dye} {et~al.}(2018){Dye}, {Lawrence}, {Read}, {Fan}, {Kerr},
  {Varricatt}, {Furnell}, {Edge}, {Irwin}, {Hambly}, {Lucas}, {Almaini},
  {Chambers}, {Green}, {Hewett}, {Liu}, {McGreer}, {Best}, {Zhang}, {Sutorius},
  {Froebrich}, {Magnier}, {Hasinger}, {Lederer}, {Bold}, \& {Tedds}}]{Dye18}
{Dye}, S., {Lawrence}, A., {Read}, M.~A., {et~al.} 2018, \mnras, 473, 5113

\bibitem[{{Euclid Collaboration} {et~al.}(2025){Euclid Collaboration}, {Fu},
  {Bouwens}, {Caputi}, {Vergani}, {Scialpi}, {Margalef-Bentabol}, {Wang},
  {Bolzonella}, {Banerji}, {Ba{\~n}ados}, {Feltre}, {Toba}, {Calhau},
  {Tarsitano}, {Cunha}, {Humphrey}, {Vietri}, {Mannucci}, {Bisogni}, {Ricci},
  {Landt}, {Spinoglio}, {Matamoro Zatarain}, {Stern}, {Page}, {Alexander},
  {Zamorani}, {Roster}, {Salvato}, {Copin}, {Sorce}, {Scott}, {Zhang}, {Lusso},
  {Wolf}, {Yang}, {Rottgering}, {Laloux}, {Siudek}, {Belladitta}, {Liu},
  {Allevato}, {Andreon}, {Auricchio}, {Baccigalupi}, {Baldi}, {Balestra},
  {Bardelli}, {Battaglia}, {Biviano}, {Branchini}, {Brescia}, {Brinchmann},
  {Camera}, {Ca{\~n}as-Herrera}, {Capobianco}, {Carbone}, {Carretero}, {Casas},
  {Castellano}, {Castignani}, {Cavuoti}, {Chambers}, {Cimatti},
  {Colodro-Conde}, {Congedo}, {Conselice}, {Conversi}, {Costille}, {Courbin},
  {Courtois}, {Cropper}, {Da Silva}, {Degaudenzi}, {De Lucia}, {Dolding},
  {Dole}, {Dubath}, {Duncan}, {Dupac}, {Dusini}, {Escoffier}, {Fabricius},
  {Farina}, {Farinelli}, {Ferriol}, {Finelli}, {Fosalba}, {Fourmanoit},
  {Frailis}, {Franceschi}, {Franzetti}, {Fumana}, {Galeotta}, {George},
  {Gillard}, {Gillis}, {Giocoli}, {Gracia-Carpio}, {Grazian}, {Grupp}, {Guzzo},
  {Haugan}, {Hoekstra}, {Holmes}, {Hook}, {Hormuth}, {Hornstrup}, {Jahnke},
  {Jhabvala}, {Joachimi}, {Keih{\"a}nen}, {Kermiche}, {Kiessling}, {Kubik},
  {K{\"u}mmel}, {Kunz}, {Kurki-Suonio}, {Laureijs}, {Le Brun}, {Ligori},
  {Lilje}, {Lindholm}, {Lloro}, {Mainetti}, {Maino}, {Maiorano}, {Mansutti},
  {Marcin}, {Marggraf}, {Markovic}, {Martinelli}, {Martinet}, {Marulli},
  {Massey}, {Medinaceli}, {Mei}, {Melchior}, {Mellier}, {Meneghetti}, {Merlin},
  {Meylan}, {Mora}, {Moresco}, {Moscardini}, {Nakajima}, {Neissner}, {Nichol},
  {Niemi}, {Padilla}, {Paltani}, {Pasian}, {Pedersen}, {Percival}, {Pettorino},
  {Pires}, {Polenta}, {Poncet}, {Popa}, {Pozzetti}, {Raison}, {Rebolo},
  {Renzi}, {Rhodes}, {Riccio}, {Romelli}, {Roncarelli}, {Rossetti}, {Saglia},
  {Sakr}, {Sapone}, {Sartoris}, {Schirmer}, {Schneider}, {Schrabback},
  {Scodeggio}, {Secroun}, {Sefusatti}, {Seidel}, {Serrano}, {Simon},
  {Sirignano}, {Sirri}, {Stanco}, {Starck}, {Steinwagner}, {Surace},
  {Tallada-Cresp{\'\i}}, {Tavagnacco}, {Taylor}, {Teplitz}, {Tereno},
  {Tessore}, {Toft}, {Toledo-Moreo}, {Torradeflot}, {Tutusaus}, {Valenziano},
  \& {Valiviita}}]{Fu26}
{Euclid Collaboration}, {Fu}, Y., {Bouwens}, R., {et~al.} 2025, arXiv e-prints,
  arXiv:2512.08803

\bibitem[{{Fan} {et~al.}(2023){Fan}, {Ba{\~n}ados}, \& {Simcoe}}]{fan23}
{Fan}, X., {Ba{\~n}ados}, E., \& {Simcoe}, R.~A. 2023, \araa, 61, 373

\bibitem[{{Fan} {et~al.}(2006){Fan}, {Strauss}, {Becker}, {White}, {Gunn},
  {Knapp}, {Richards}, {Schneider}, {Brinkmann}, \& {Fukugita}}]{Fan06}
{Fan}, X., {Strauss}, M.~A., {Becker}, R.~H., {et~al.} 2006, \aj, 132, 117

\bibitem[{{Fan} {et~al.}(2019){Fan}, {Wang}, {Yang}, {Keeton}, {Yue},
  {Zabludoff}, {Bian}, {Bonaglia}, {Georgiev}, {Hennawi}, {Li}, {McGreer},
  {Naidu}, {Pacucci}, {Rabien}, {Thompson}, {Venemans}, {Walter}, {Wang}, \&
  {Wu}}]{Fan19}
{Fan}, X., {Wang}, F., {Yang}, J., {et~al.} 2019, \apjl, 870, L11

\bibitem[{{Feder} {et~al.}(2024){Feder}, {Masters}, {Lee}, {Bock}, {Chiang},
  {Choi}, {Dor{\'e}}, {Hemmati}, \& {Ilbert}}]{Feder24}
{Feder}, R.~M., {Masters}, D.~C., {Lee}, B., {et~al.} 2024, \apj, 972, 68

\bibitem[{{Fitzpatrick}(1999)}]{Fitzpatrick99}
{Fitzpatrick}, E.~L. 1999, \pasp, 111, 63

\bibitem[{{Gaia Collaboration} {et~al.}(2016){Gaia Collaboration}, {Prusti},
  {de Bruijne}, {Brown}, {Vallenari}, {Babusiaux}, {Bailer-Jones}, {Bastian},
  {Biermann}, {Evans}, {Eyer}, {Jansen}, {Jordi}, {Klioner}, {Lammers},
  {Lindegren}, {Luri}, {Mignard}, {Milligan}, {Panem}, {Poinsignon},
  {Pourbaix}, {Randich}, {Sarri}, {Sartoretti}, {Siddiqui}, {Soubiran},
  {Valette}, {van Leeuwen}, {Walton}, {Aerts}, {Arenou}, {Cropper}, {Drimmel},
  {H{\o}g}, {Katz}, {Lattanzi}, {O'Mullane}, {Grebel}, {Holland}, {Huc},
  {Passot}, {Bramante}, {Cacciari}, {Casta{\~n}eda}, {Chaoul}, {Cheek}, {De
  Angeli}, {Fabricius}, {Guerra}, {Hern{\'a}ndez}, {Jean-Antoine-Piccolo},
  {Masana}, {Messineo}, {Mowlavi}, {Nienartowicz}, {Ord{\'o}{\~n}ez-Blanco},
  {Panuzzo}, {Portell}, {Richards}, {Riello}, {Seabroke}, {Tanga},
  {Th{\'e}venin}, {Torra}, {Els}, {Gracia-Abril}, {Comoretto},
  {Garcia-Reinaldos}, {Lock}, {Mercier}, {Altmann}, {Andrae}, {Astraatmadja},
  {Bellas-Velidis}, {Benson}, {Berthier}, {Blomme}, {Busso}, {Carry},
  {Cellino}, {Clementini}, {Cowell}, {Creevey}, {Cuypers}, {Davidson}, {De
  Ridder}, {de Torres}, {Delchambre}, {Dell'Oro}, {Ducourant}, {Fr{\'e}mat},
  {Garc{\'\i}a-Torres}, {Gosset}, {Halbwachs}, {Hambly}, {Harrison}, {Hauser},
  {Hestroffer}, {Hodgkin}, {Huckle}, {Hutton}, {Jasniewicz}, {Jordan},
  {Kontizas}, {Korn}, {Lanzafame}, {Manteiga}, {Moitinho}, {Muinonen},
  {Osinde}, {Pancino}, {Pauwels}, {Petit}, {Recio-Blanco}, {Robin}, {Sarro},
  {Siopis}, {Smith}, {Smith}, {Sozzetti}, {Thuillot}, {van Reeven}, {Viala},
  {Abbas}, {Abreu Aramburu}, {Accart}, {Aguado}, {Allan}, {Allasia},
  {Altavilla}, {{\'A}lvarez}, {Alves}, {Anderson}, {Andrei}, {Anglada Varela},
  {Antiche}, {Antoja}, {Ant{\'o}n}, {Arcay}, {Atzei}, {Ayache}, {Bach},
  {Baker}, {Balaguer-N{\'u}{\~n}ez}, {Barache}, {Barata}, {Barbier}, {Barblan},
  {Baroni}, {Barrado y Navascu{\'e}s}, {Barros}, {Barstow}, {Becciani},
  {Bellazzini}, {Bellei}, {Bello Garc{\'\i}a}, {Belokurov}, {Bendjoya},
  {Berihuete}, {Bianchi}, {Bienaym{\'e}}, {Billebaud}, {Blagorodnova},
  {Blanco-Cuaresma}, {Boch}, {Bombrun}, {Borrachero}, {Bouquillon}, {Bourda},
  {Bouy}, {Bragaglia}, {Breddels}, {Brouillet}, {Br{\"u}semeister},
  {Bucciarelli}, {Budnik}, {Burgess}, {Burgon}, {Burlacu}, {Busonero}, {Buzzi},
  {Caffau}, {Cambras}, {Campbell}, {Cancelliere}, {Cantat-Gaudin}, {Carlucci},
  {Carrasco}, {Castellani}, {Charlot}, {Charnas}, {Charvet}, {Chassat},
  {Chiavassa}, {Clotet}, {Cocozza}, {Collins}, {Collins}, \& {Costigan}}]{Gaia}
{Gaia Collaboration}, {Prusti}, T., {de Bruijne}, J.~H.~J., {et~al.} 2016,
  \aap, 595, A1

\bibitem[{{Gaia Collaboration} {et~al.}(2023){Gaia Collaboration}, {Vallenari},
  {Brown}, {Prusti}, {de Bruijne}, {Arenou}, {Babusiaux}, {Biermann},
  {Creevey}, {Ducourant}, {Evans}, {Eyer}, {Guerra}, {Hutton}, {Jordi},
  {Klioner}, {Lammers}, {Lindegren}, {Luri}, {Mignard}, {Panem}, {Pourbaix},
  {Randich}, {Sartoretti}, {Soubiran}, {Tanga}, {Walton}, {Bailer-Jones},
  {Bastian}, {Drimmel}, {Jansen}, {Katz}, {Lattanzi}, {van Leeuwen}, {Bakker},
  {Cacciari}, {Casta{\~n}eda}, {De Angeli}, {Fabricius}, {Fouesneau},
  {Fr{\'e}mat}, {Galluccio}, {Guerrier}, {Heiter}, {Masana}, {Messineo},
  {Mowlavi}, {Nicolas}, {Nienartowicz}, {Pailler}, {Panuzzo}, {Riclet}, {Roux},
  {Seabroke}, {Sordo}, {Th{\'e}venin}, {Gracia-Abril}, {Portell}, {Teyssier},
  {Altmann}, {Andrae}, {Audard}, {Bellas-Velidis}, {Benson}, {Berthier},
  {Blomme}, {Burgess}, {Busonero}, {Busso}, {C{\'a}novas}, {Carry}, {Cellino},
  {Cheek}, {Clementini}, {Damerdji}, {Davidson}, {de Teodoro}, {Nu{\~n}ez
  Campos}, {Delchambre}, {Dell'Oro}, {Esquej}, {Fern{\'a}ndez-Hern{\'a}ndez},
  {Fraile}, {Garabato}, {Garc{\'\i}a-Lario}, {Gosset}, {Haigron}, {Halbwachs},
  {Hambly}, {Harrison}, {Hern{\'a}ndez}, {Hestroffer}, {Hodgkin}, {Holl},
  {Jan{\ss}en}, {Jevardat de Fombelle}, {Jordan}, {Krone-Martins}, {Lanzafame},
  {L{\"o}ffler}, {Marchal}, {Marrese}, {Moitinho}, {Muinonen}, {Osborne},
  {Pancino}, {Pauwels}, {Recio-Blanco}, {Reyl{\'e}}, {Riello}, {Rimoldini},
  {Roegiers}, {Rybizki}, {Sarro}, {Siopis}, {Smith}, {Sozzetti}, {Utrilla},
  {van Leeuwen}, {Abbas}, {{\'A}brah{\'a}m}, {Abreu Aramburu}, {Aerts},
  {Aguado}, {Ajaj}, {Aldea-Montero}, {Altavilla}, {{\'A}lvarez}, {Alves},
  {Anders}, {Anderson}, {Anglada Varela}, {Antoja}, {Baines}, {Baker},
  {Balaguer-N{\'u}{\~n}ez}, {Balbinot}, {Balog}, {Barache}, {Barbato},
  {Barros}, {Barstow}, {Bartolom{\'e}}, {Bassilana}, {Bauchet}, {Becciani},
  {Bellazzini}, {Berihuete}, {Bernet}, {Bertone}, {Bianchi}, {Binnenfeld},
  {Blanco-Cuaresma}, {Blazere}, {Boch}, {Bombrun}, {Bossini}, {Bouquillon},
  {Bragaglia}, {Bramante}, {Breedt}, {Bressan}, {Brouillet}, {Brugaletta},
  {Bucciarelli}, {Burlacu}, {Butkevich}, {Buzzi}, {Caffau}, {Cancelliere},
  {Cantat-Gaudin}, {Carballo}, {Carlucci}, {Carnerero}, {Carrasco},
  {Casamiquela}, {Castellani}, {Castro-Ginard}, {Chaoul}, {Charlot}, {Chemin},
  {Chiaramida}, {Chiavassa}, {Chornay}, {Comoretto}, {Contursi}, {Cooper},
  {Cornez}, {Cowell}, {Crifo}, {Cropper}, {Crosta}, {Crowley}, {Dafonte},
  {Dapergolas}, {David}, {David}, {de Laverny}, {De Luise}, \& {De
  March}}]{GaiaDR3}
{Gaia Collaboration}, {Vallenari}, A., {Brown}, A.~G.~A., {et~al.} 2023, \aap,
  674, A1

\bibitem[{{Giallongo} {et~al.}(2015){Giallongo}, {Grazian}, {Fiore}, {Fontana},
  {Pentericci}, {Vanzella}, {Dickinson}, {Kocevski}, {Castellano}, {Cristiani},
  {Ferguson}, {Finkelstein}, {Grogin}, {Hathi}, {Koekemoer}, {Newman}, \&
  {Salvato}}]{Giallongo15}
{Giallongo}, E., {Grazian}, A., {Fiore}, F., {et~al.} 2015, \aap, 578, A83

\bibitem[{{Gordon} {et~al.}(2003){Gordon}, {Clayton}, {Misselt}, {Landolt}, \&
  {Wolff}}]{Gordon03}
{Gordon}, K.~D., {Clayton}, G.~C., {Misselt}, K.~A., {Landolt}, A.~U., \&
  {Wolff}, M.~J. 2003, \apj, 594, 279

\bibitem[{{Guarneri} {et~al.}(2022){Guarneri}, {Calderone}, {Cristiani},
  {Porru}, {Fontanot}, {Boutsia}, {Cupani}, {Grazian}, {D'Odorico}, {Murphy},
  {Bongiorno}, {Saccheo}, \& {Nicastro}}]{Guarneri22}
{Guarneri}, F., {Calderone}, G., {Cristiani}, S., {et~al.} 2022, \mnras, 517,
  2436

\bibitem[{{Haardt} \& {Madau}(2012)}]{HM12}
{Haardt}, F. \& {Madau}, P. 2012, \apj, 746, 125

\bibitem[{{Hern{\'a}n-Caballero} {et~al.}(2016){Hern{\'a}n-Caballero},
  {Hatziminaoglou}, {Alonso-Herrero}, \& {Mateos}}]{Hernan16}
{Hern{\'a}n-Caballero}, A., {Hatziminaoglou}, E., {Alonso-Herrero}, A., \&
  {Mateos}, S. 2016, \mnras, 463, 2064

\bibitem[{{Inayoshi} {et~al.}(2020){Inayoshi}, {Visbal}, \&
  {Haiman}}]{Inayoshi20}
{Inayoshi}, K., {Visbal}, E., \& {Haiman}, Z. 2020, \araa, 58, 27

\bibitem[{{Jiang} {et~al.}(2018){Jiang}, {Hu}, {Xu}, {Dai}, {Zhang}, {Wang}, \&
  {Chen}}]{NGPS}
{Jiang}, H., {Hu}, Z., {Xu}, M., {et~al.} 2018, in Society of Photo-Optical
  Instrumentation Engineers (SPIE) Conference Series, Vol. 10702, Ground-based
  and Airborne Instrumentation for Astronomy VII, ed. C.~J. {Evans},
  L.~{Simard}, \& H.~{Takami}, 107022L

\bibitem[{{Kulkarni} {et~al.}(2019){Kulkarni}, {Worseck}, \&
  {Hennawi}}]{Kulkarni19qlf}
{Kulkarni}, G., {Worseck}, G., \& {Hennawi}, J.~F. 2019, \mnras, 488, 1035

\bibitem[{{Lang} {et~al.}(2016){Lang}, {Hogg}, \& {Mykytyn}}]{Tractor}
{Lang}, D., {Hogg}, D.~W., \& {Mykytyn}, D. 2016, {The Tractor: Probabilistic
  astronomical source detection and measurement}, Astrophysics Source Code
  Library, record ascl:1604.008

\bibitem[{{Lapi} {et~al.}(2006){Lapi}, {Shankar}, {Mao}, {Granato}, {Silva},
  {De Zotti}, \& {Danese}}]{Lapi06}
{Lapi}, A., {Shankar}, F., {Mao}, J., {et~al.} 2006, \apj, 650, 42

\bibitem[{{Lyke} {et~al.}(2020){Lyke}, {Higley}, {McLane}, {Schurhammer},
  {Myers}, {Ross}, {Dawson}, {Chabanier}, {Martini}, {Busca}, {Mas des
  Bourboux}, {Salvato}, {Streblyanska}, {Zarrouk}, {Burtin}, {Anderson},
  {Bautista}, {Bizyaev}, {Brandt}, {Brinkmann}, {Brownstein}, {Comparat},
  {Green}, {de la Macorra}, {Mu{\~n}oz Guti{\'e}rrez}, {Hou}, {Newman},
  {Palanque-Delabrouille}, {P{\^a}ris}, {Percival}, {Petitjean}, {Rich},
  {Rossi}, {Schneider}, {Smith}, {Vivek}, \& {Weaver}}]{Lyke20}
{Lyke}, B.~W., {Higley}, A.~N., {McLane}, J.~N., {et~al.} 2020, \apjs, 250, 8

\bibitem[{{Mainzer} {et~al.}(2011){Mainzer}, {Bauer}, {Grav}, {Masiero},
  {Cutri}, {Dailey}, {Eisenhardt}, {McMillan}, {Wright}, {Walker}, {Jedicke},
  {Spahr}, {Tholen}, {Alles}, {Beck}, {Brandenburg}, {Conrow}, {Evans},
  {Fowler}, {Jarrett}, {Marsh}, {Masci}, {McCallon}, {Wheelock}, {Wittman},
  {Wyatt}, {DeBaun}, {Elliott}, {Elsbury}, {Gautier}, {Gomillion}, {Leisawitz},
  {Maleszewski}, {Micheli}, \& {Wilkins}}]{NeoWISE}
{Mainzer}, A., {Bauer}, J., {Grav}, T., {et~al.} 2011, \apj, 731, 53

\bibitem[{{Marocco} {et~al.}(2021){Marocco}, {Eisenhardt}, {Fowler},
  {Kirkpatrick}, {Meisner}, {Schlafly}, {Stanford}, {Garcia}, {Caselden},
  {Cushing}, {Cutri}, {Faherty}, {Gelino}, {Gonzalez}, {Jarrett}, {Koontz},
  {Mainzer}, {Marchese}, {Mobasher}, {Schlegel}, {Stern}, {Teplitz}, \&
  {Wright}}]{CatWISE2020}
{Marocco}, F., {Eisenhardt}, P. R.~M., {Fowler}, J.~W., {et~al.} 2021, \apjs,
  253, 8

\bibitem[{{Martin} {et~al.}(2010){Martin}, {Moore}, {Morrissey}, {Matuszewski},
  {Rahman}, {Adkins}, \& {Epps}}]{KCWI}
{Martin}, C., {Moore}, A., {Morrissey}, P., {et~al.} 2010, in Society of
  Photo-Optical Instrumentation Engineers (SPIE) Conference Series, Vol. 7735,
  Ground-based and Airborne Instrumentation for Astronomy III, ed. I.~S.
  {McLean}, S.~K. {Ramsay}, \& H.~{Takami}, 77350M

\bibitem[{{Martins}(2017)}]{Martins17}
{Martins}, C.~J.~A.~P. 2017, Reports on Progress in Physics, 80, 126902

\bibitem[{{Marulli} {et~al.}(2008){Marulli}, {Bonoli}, {Branchini},
  {Moscardini}, \& {Springel}}]{Marulli08}
{Marulli}, F., {Bonoli}, S., {Branchini}, E., {Moscardini}, L., \& {Springel},
  V. 2008, \mnras, 385, 1846

\bibitem[{{Nanni} {et~al.}(2022){Nanni}, {Hennawi}, {Wang}, {Yang},
  {Schindler}, \& {Fan}}]{Nanni22}
{Nanni}, R., {Hennawi}, J.~F., {Wang}, F., {et~al.} 2022, \mnras, 515, 3224

\bibitem[{{Oke} \& {Gunn}(1983)}]{Oke83}
{Oke}, J.~B. \& {Gunn}, J.~E. 1983, \apj, 266, 713

\bibitem[{{Onken} {et~al.}(2024){Onken}, {Wolf}, {Bessell}, {Chang}, {Luvaul},
  {Tonry}, {White}, \& {Da Costa}}]{Onken24}
{Onken}, C.~A., {Wolf}, C., {Bessell}, M.~S., {et~al.} 2024, \pasa, 41, e061

\bibitem[{{Onken} {et~al.}(2022){Onken}, {Wolf}, {Bian}, {Fan}, {Hon},
  {Raithel}, {Tisserand}, \& {Lai}}]{Onken22}
{Onken}, C.~A., {Wolf}, C., {Bian}, F., {et~al.} 2022, \mnras, 511, 572

\bibitem[{{Planck Collaboration} {et~al.}(2020){Planck Collaboration},
  {Aghanim}, {Akrami}, {Ashdown}, {Aumont}, {Baccigalupi}, {Ballardini},
  {Banday}, {Barreiro}, {Bartolo}, {Basak}, {Battye}, {Benabed}, {Bernard},
  {Bersanelli}, {Bielewicz}, {Bock}, {Bond}, {Borrill}, {Bouchet}, {Boulanger},
  {Bucher}, {Burigana}, {Butler}, {Calabrese}, {Cardoso}, {Carron},
  {Challinor}, {Chiang}, {Chluba}, {Colombo}, {Combet}, {Contreras}, {Crill},
  {Cuttaia}, {de Bernardis}, {de Zotti}, {Delabrouille}, {Delouis}, {Di
  Valentino}, {Diego}, {Dor{\'e}}, {Douspis}, {Ducout}, {Dupac}, {Dusini},
  {Efstathiou}, {Elsner}, {En{\ss}lin}, {Eriksen}, {Fantaye}, {Farhang},
  {Fergusson}, {Fernandez-Cobos}, {Finelli}, {Forastieri}, {Frailis},
  {Fraisse}, {Franceschi}, {Frolov}, {Galeotta}, {Galli}, {Ganga},
  {G{\'e}nova-Santos}, {Gerbino}, {Ghosh}, {Gonz{\'a}lez-Nuevo}, {G{\'o}rski},
  {Gratton}, {Gruppuso}, {Gudmundsson}, {Hamann}, {Handley}, {Hansen},
  {Herranz}, {Hildebrandt}, {Hivon}, {Huang}, {Jaffe}, {Jones}, {Karakci},
  {Keih{\"a}nen}, {Keskitalo}, {Kiiveri}, {Kim}, {Kisner}, {Knox},
  {Krachmalnicoff}, {Kunz}, {Kurki-Suonio}, {Lagache}, {Lamarre}, {Lasenby},
  {Lattanzi}, {Lawrence}, {Le Jeune}, {Lemos}, {Lesgourgues}, {Levrier},
  {Lewis}, {Liguori}, {Lilje}, {Lilley}, {Lindholm}, {L{\'o}pez-Caniego},
  {Lubin}, {Ma}, {Mac{\'\i}as-P{\'e}rez}, {Maggio}, {Maino}, {Mandolesi},
  {Mangilli}, {Marcos-Caballero}, {Maris}, {Martin}, {Martinelli},
  {Mart{\'\i}nez-Gonz{\'a}lez}, {Matarrese}, {Mauri}, {McEwen}, {Meinhold},
  {Melchiorri}, {Mennella}, {Migliaccio}, {Millea}, {Mitra},
  {Miville-Desch{\^e}nes}, {Molinari}, {Montier}, {Morgante}, {Moss}, {Natoli},
  {N{\o}rgaard-Nielsen}, {Pagano}, {Paoletti}, {Partridge}, {Patanchon},
  {Peiris}, {Perrotta}, {Pettorino}, {Piacentini}, {Polastri}, {Polenta},
  {Puget}, {Rachen}, {Reinecke}, {Remazeilles}, {Renzi}, {Rocha}, {Rosset},
  {Roudier}, {Rubi{\~n}o-Mart{\'\i}n}, {Ruiz-Granados}, {Salvati}, {Sandri},
  {Savelainen}, {Scott}, {Shellard}, {Sirignano}, {Sirri}, {Spencer},
  {Sunyaev}, {Suur-Uski}, {Tauber}, {Tavagnacco}, {Tenti}, {Toffolatti},
  {Tomasi}, {Trombetti}, {Valenziano}, {Valiviita}, {Van Tent}, {Vibert},
  {Vielva}, {Villa}, {Vittorio}, {Wandelt}, {Wehus}, {White}, {White},
  {Zacchei}, \& {Zonca}}]{Planck18}
{Planck Collaboration}, {Aghanim}, N., {Akrami}, Y., {et~al.} 2020, \aap, 641,
  A6

\bibitem[{{Rafelski} {et~al.}(2012){Rafelski}, {Wolfe}, {Prochaska},
  {Neeleman}, \& {Mendez}}]{Rafelski12}
{Rafelski}, M., {Wolfe}, A.~M., {Prochaska}, J.~X., {Neeleman}, M., \&
  {Mendez}, A.~J. 2012, \apj, 755, 89

\bibitem[{{Riello} {et~al.}(2021){Riello}, {De Angeli}, {Evans}, {Montegriffo},
  {Carrasco}, {Busso}, {Palaversa}, {Burgess}, {Diener}, {Davidson}, {Rowell},
  {Fabricius}, {Jordi}, {Bellazzini}, {Pancino}, {Harrison}, {Cacciari}, {van
  Leeuwen}, {Hambly}, {Hodgkin}, {Osborne}, {Altavilla}, {Barstow}, {Brown},
  {Castellani}, {Cowell}, {De Luise}, {Gilmore}, {Giuffrida}, {Hidalgo},
  {Holland}, {Marinoni}, {Pagani}, {Piersimoni}, {Pulone}, {Ragaini}, {Rainer},
  {Richards}, {Sanna}, {Walton}, {Weiler}, \& {Yoldas}}]{Riello21}
{Riello}, M., {De Angeli}, F., {Evans}, D.~W., {et~al.} 2021, \aap, 649, A3

\bibitem[{{Schlafly} \& {Finkbeiner}(2011)}]{SB11}
{Schlafly}, E.~F. \& {Finkbeiner}, D.~P. 2011, \apj, 737, 103

\bibitem[{{Selsing} {et~al.}(2016){Selsing}, {Fynbo}, {Christensen}, \&
  {Krogager}}]{Selsing16}
{Selsing}, J., {Fynbo}, J.~P.~U., {Christensen}, L., \& {Krogager}, J.-K. 2016,
  \aap, 585, A87

\bibitem[{{Silverman} {et~al.}(2008){Silverman}, {Green}, {Barkhouse}, {Kim},
  {Kim}, {Wilkes}, {Cameron}, {Hasinger}, {Jannuzi}, {Smith}, {Smith}, \&
  {Tananbaum}}]{Silverman08}
{Silverman}, J.~D., {Green}, P.~J., {Barkhouse}, W.~A., {et~al.} 2008, \apj,
  679, 118

\bibitem[{{Soltan}(1982)}]{Soltan82}
{Soltan}, A. 1982, \mnras, 200, 115

\bibitem[{{Storey-Fisher} {et~al.}(2024){Storey-Fisher}, {Hogg}, {Rix},
  {Eilers}, {Fabbian}, {Blanton}, \& {Alonso}}]{Storey-Fisher24}
{Storey-Fisher}, K., {Hogg}, D.~W., {Rix}, H.-W., {et~al.} 2024, \apj, 964, 69

\bibitem[{{Volonteri}(2010)}]{Volonteri10}
{Volonteri}, M. 2010, \aapr, 18, 279

\bibitem[{{Wang} {et~al.}(2015){Wang}, {Wu}, {Fan}, {Yang}, {Cai}, {Yi}, {Zuo},
  {Wang}, {McGreer}, {Ho}, {Kim}, {Yang}, {Bian}, \& {Jiang}}]{Wang15}
{Wang}, F., {Wu}, X.-B., {Fan}, X., {et~al.} 2015, \apjl, 807, L9

\bibitem[{{Wang} {et~al.}(2016){Wang}, {Wu}, {Fan}, {Yang}, {Yi}, {Bian},
  {McGreer}, {Yang}, {Ai}, {Dong}, {Zuo}, {Jiang}, {Green}, {Wang}, {Cai},
  {Wang}, \& {Yue}}]{Wang16}
{Wang}, F., {Wu}, X.-B., {Fan}, X., {et~al.} 2016, \apj, 819, 24

\bibitem[{{Wilczynska} {et~al.}(2020){Wilczynska}, {Webb}, {Bainbridge},
  {Barrow}, {Bosman}, {Carswell}, {D{\k{a}}browski}, {Dumont}, {Lee}, {Leite},
  {Leszczy{\'n}ska}, {Liske}, {Marosek}, {Martins}, {Milakovi{\'c}}, {Molaro},
  \& {Pasquini}}]{Wilczynska20}
{Wilczynska}, M.~R., {Webb}, J.~K., {Bainbridge}, M., {et~al.} 2020, Science
  Advances, 6, eaay9672

\bibitem[{{Wolf} {et~al.}(2020){Wolf}, {Hon}, {Bian}, {Onken}, {Alonzi},
  {Bessell}, {Li}, {Schmidt}, \& {Tisserand}}]{Wolf20}
{Wolf}, C., {Hon}, W.~J., {Bian}, F., {et~al.} 2020, \mnras, 491, 1970

\bibitem[{{Wright} {et~al.}(2010){Wright}, {Eisenhardt}, {Mainzer}, {Ressler},
  {Cutri}, {Jarrett}, {Kirkpatrick}, {Padgett}, {McMillan}, {Skrutskie},
  {Stanford}, {Cohen}, {Walker}, {Mather}, {Leisawitz}, {Gautier}, {McLean},
  {Benford}, {Lonsdale}, {Blain}, {Mendez}, {Irace}, {Duval}, {Liu}, {Royer},
  {Heinrichsen}, {Howard}, {Shannon}, {Kendall}, {Walsh}, {Larsen}, {Cardon},
  {Schick}, {Schwalm}, {Abid}, {Fabinsky}, {Naes}, \& {Tsai}}]{WISE}
{Wright}, E.~L., {Eisenhardt}, P. R.~M., {Mainzer}, A.~K., {et~al.} 2010, \aj,
  140, 1868

\bibitem[{{Wu} {et~al.}(2015){Wu}, {Wang}, {Fan}, {Yi}, {Zuo}, {Bian}, {Jiang},
  {McGreer}, {Wang}, {Yang}, {Yang}, {Thompson}, \& {Beletsky}}]{Wu15}
{Wu}, X.-B., {Wang}, F., {Fan}, X., {et~al.} 2015, \nat, 518, 512

\bibitem[{{Yang} {et~al.}(2017){Yang}, {Fan}, {Wu}, {Wang}, {Bian}, {Yang},
  {McGreer}, {Yi}, {Jiang}, {Green}, {Yue}, {Wang}, {Li}, {Ding}, {Dye}, \&
  {Lawrence}}]{Yang17}
{Yang}, J., {Fan}, X., {Wu}, X.-B., {et~al.} 2017, \aj, 153, 184

\bibitem[{{Yang} {et~al.}(2019){Yang}, {Wang}, {Fan}, {Wu}, {Bian},
  {Ba{\~n}ados}, {Yue}, {Schindler}, {Yang}, {Jiang}, {McGreer}, {Green}, \&
  {Dye}}]{Yang19b}
{Yang}, J., {Wang}, F., {Fan}, X., {et~al.} 2019, \apj, 871, 199

\end{thebibliography}

 \newcommand{\noop}[1]{}

\begin{appendix}

\section{Table of new $z>4$ quasars}

\begin{table*}[]
    \centering
    \begin{tabular}{lccclcccc}
        Name & RA & Dec & $z$ & Selection & $G_{\rm RP}$ & $W1$ & $W2$ & $M_{1450}$ \\
        \hline \hline 
        SPX J0003$-$1631${^a}$${^b}$ & 00:03:25.50 & $-$16:31:09.09 & 5.22 & fiducial & 19.18$\pm$0.01 & 16.02$\pm$0.03 & 15.23$\pm$0.04 & -26.7 \\
        SPX J0021$+$3520${^b}$ & 00:21:01.74 & $+$35:20:43.26 & 4.49 & fiducial & 18.32$\pm$0.01 & 14.52$\pm$0.02 & 14.06$\pm$0.01 & -28.5 \\
        SPX J0033$-$7109 & 00:33:24.72 & $-$71:09:31.67 & 5.08 & fiducial & 19.01$\pm$0.05 & 15.59$\pm$0.02 & 15.06$\pm$0.02 & -26.8 \\
        SPX J0113$+$1542 & 01:13:59.62 & $+$15:42:22.38 & 4.40 & $G_{\rm RP}$ faint & 19.26$\pm$0.02 & 15.47$\pm$0.02 & 14.81$\pm$0.03 & -26.5 \\
        SPX J0130$-$5414${^a}$ & 01:30:08.14 & $-$54:14:38.59 & 4.69 & fiducial & 18.72$\pm$0.02 & 15.27$\pm$0.02 & 14.79$\pm$0.02 & -26.6 \\
        SPX J0137$-$5226${^a}$ & 01:37:25.16 & $-$52:26:29.45 & 5.06 & $G_{\rm BP}$ flux & 19.25$\pm$0.01 & 15.91$\pm$0.02 & 15.27$\pm$0.03 & -26.6 \\
        SPX J0149$-$6442${^a}$ & 01:49:04.23 & $-$64:42:11.81 & 4.60 & fiducial & 18.50$\pm$0.02 & 15.63$\pm$0.02 & 15.25$\pm$0.03 & -27.6 \\
        SPX J0154$-$2347${^a}$ & 01:54:11.99 & $-$23:47:34.45 & 4.77 & fiducial & 18.54$\pm$0.01 & 15.28$\pm$0.02 & 14.78$\pm$0.03 & -27.0 \\
        SPX J0159$-$7235 & 01:59:24.70 & $-$72:35:30.08 & 4.42 & fiducial & 18.89$\pm$0.03 & 16.01$\pm$0.03 & 15.62$\pm$0.03 & -27.3 \\
        SPX J0159$-$3919${^a}$ & 01:59:38.89 & $-$39:19:41.54 & 5.15 & fiducial & 18.91$\pm$0.03 & 15.80$\pm$0.02 & 15.22$\pm$0.03 & -27.4 \\
        SPX J0200$-$2942${^b}$ & 02:00:52.99 & $-$29:42:49.15 & 5.32 & $G_{\rm BP}$ flux & 19.42$\pm$0.02 & 16.25$\pm$0.03 & 15.32$\pm$0.03 & -27.2 \\
        SPX J0206$-$3444 & 02:06:26.33 & $-$34:44:10.28 & 4.76 & fiducial & 19.08$\pm$0.03 & 15.65$\pm$0.02 & 15.26$\pm$0.03 & -26.6 \\
        SPX J0222$-$5423 & 02:22:26.26 & $-$54:23:58.29 & 4.03 & fiducial & 19.09$\pm$0.04 & 14.45$\pm$0.01 & 13.96$\pm$0.01 & -26.3 \\
        SPX J0223$-$2433${^a}$ & 02:23:40.06 & $-$24:33:37.44 & 5.06 & fiducial & 18.64$\pm$0.01 & 15.42$\pm$0.03 & 14.93$\pm$0.04 & -27.7 \\
        SPX J0225$-$5450${^a}$ & 02:25:26.96 & $-$54:50:21.28 & 4.82 & fiducial & 18.74$\pm$0.05 & 15.88$\pm$0.02 & 15.32$\pm$0.03 & -27.7 \\
        SPX J0240$+$3950${^b}$ & 02:40:55.52 & $+$39:50:35.29 & 4.70 & fiducial & 18.10$\pm$0.04 & 14.80$\pm$0.02 & 14.38$\pm$0.02 & -28.2 \\
        SPX J0308$-$4841 & 03:08:34.69 & $-$48:41:28.16 & 4.30 & fiducial & 18.19$\pm$0.01 & 13.96$\pm$0.05 & 13.49$\pm$0.04 & -28.6 \\
        SPX J0324$+$2844 & 03:24:58.59 & $+$28:44:45.25 & 4.35 & fiducial & 18.93$\pm$0.02 & 15.32$\pm$0.02 & 14.97$\pm$0.03 & -27.2 \\
        SPX J0328$-$1748${^b}$ & 03:28:30.16 & $-$17:48:26.45 & 4.30 & fiducial & 18.98$\pm$0.03 & 15.79$\pm$0.02 & 15.31$\pm$0.04 & -27.2 \\
        SPX J0340$-$0939 & 03:40:45.84 & $-$09:39:30.31 & 4.81 & $G_{\rm BP}$ flux & 19.16$\pm$0.02 & 15.63$\pm$0.02 & 15.00$\pm$0.03 & -26.5 \\
        SPX J0350$-$5803${^a}$ & 03:50:00.48 & $-$58:03:23.39 & 4.62 & fiducial & 18.60$\pm$0.05 & 15.74$\pm$0.02 & 15.36$\pm$0.03 & -27.5 \\
        SPX J0405$-$1335 & 04:05:02.64 & $-$13:35:56.95 & 4.21 & fiducial & 18.71$\pm$0.03 & 15.04$\pm$0.02 & 14.50$\pm$0.02 & -27.2 \\
        SPX J0413$+$0328${^b}$ & 04:13:57.24 & $+$03:28:10.11 & 5.26 & fiducial & 18.91$\pm$0.07 & 14.94$\pm$0.02 & 14.11$\pm$0.02 & -27.9 \\
        SPX J0439$-$6316${^a}$ & 04:39:27.63 & $-$63:16:56.80 & 4.93 & fiducial & 18.94$\pm$0.02 & 16.05$\pm$0.02 & 15.49$\pm$0.03 & -27.3 \\
        SPX J0533$+$5549${^b}$ & 05:33:43.62 & $+$55:49:34.91 & 4.55 & fiducial & 19.00$\pm$0.05 & 14.78$\pm$0.02 & 14.39$\pm$0.02 & -27.8 \\
        SPX J0549$-$7348 & 05:49:05.65 & $-$73:48:34.48 & 4.70 & fiducial & 18.91$\pm$0.07 & 15.33$\pm$0.02 & 14.97$\pm$0.02 & -27.3 \\
        SPX J0550$-$1900 & 05:50:24.06 & $-$19:00:28.09 & 4.54 & fiducial & 18.73$\pm$0.05 & 15.79$\pm$0.02 & 15.39$\pm$0.04 & -26.9 \\
        SPX J0553$+$7454 & 05:53:41.90 & $+$74:54:38.47 & 4.94 & fiducial & 18.79$\pm$0.02 & 15.82$\pm$0.02 & 15.28$\pm$0.03 & -27.6 \\
        SPX J0601$-$2621${^b}$ & 06:01:14.23 & $-$26:21:19.35 & 4.61 & $G_{\rm RP}$ faint & 19.40$\pm$0.03 & 16.50$\pm$0.03 & 15.90$\pm$0.05 & -26.6 \\
        SPX J0615$-$4914 & 06:15:56.91 & $-$49:14:02.51 & 4.70 & fiducial & 18.45$\pm$0.02 & 14.96$\pm$0.01 & 14.60$\pm$0.02 & -27.4 \\
        SPX J0642$+$7444 & 06:42:52.26 & $+$74:44:57.29 & 4.59 & fiducial & 18.28$\pm$0.03 & 15.03$\pm$0.02 & 14.63$\pm$0.02 & -27.8 \\
        SPX J0646$-$3541 & 06:46:11.66 & $-$35:41:07.27 & 4.29 & fiducial & 18.59$\pm$0.02 & 14.69$\pm$0.02 & 14.38$\pm$0.02 & -28.6 \\
        SPX J0650$+$7253 & 06:50:56.45 & $+$72:53:14.64 & 4.57 & fiducial & 18.01$\pm$0.02 & 14.98$\pm$0.02 & 14.57$\pm$0.02 & -27.9 \\
        SPX J0701$+$5243${^b}$ & 07:01:39.95 & $+$52:43:03.86 & 5.49 & fiducial & 18.89$\pm$0.03 & 15.42$\pm$0.02 & 14.68$\pm$0.02 & -27.9 \\
        SPX J0703$+$3506 & 07:03:34.03 & $+$35:06:11.84 & 4.51 & fiducial & 18.49$\pm$0.03 & 15.52$\pm$0.02 & 15.08$\pm$0.03 & -27.8 \\
        SPX J0709$+$4123${^b}$ & 07:09:53.51 & $+$41:23:49.80 & 4.08 & fiducial & 18.41$\pm$0.02 & 13.82$\pm$0.01 & 13.45$\pm$0.01 & -26.8 \\
        SPX J0717$+$7126${^b}$ & 07:17:46.41 & $+$71:26:10.00 & 4.83 & $G_{\rm RP}$ faint & 19.35$\pm$0.02 & 15.90$\pm$0.02 & 15.29$\pm$0.03 & -26.8 \\
        SPX J0930$-$3000${^b}$ & 09:30:07.09 & $-$30:00:32.03 & 4.58 & fiducial & 18.85$\pm$0.03 & 15.66$\pm$0.02 & 15.24$\pm$0.03 & -27.6 \\
        SPX J0933$+$7541 & 09:33:37.71 & $+$75:41:58.13 & 5.05 & fiducial & 18.96$\pm$0.05 & 15.83$\pm$0.02 & 15.32$\pm$0.03 & -27.8 \\
        SPX J1024$-$1030 & 10:24:25.59 & $-$10:30:06.45 & 4.77 & fiducial & 19.12$\pm$0.04 & 15.68$\pm$0.02 & 15.14$\pm$0.03 & -27.3 \\
        SPX J1025$-$2443${^b}$ & 10:25:25.92 & $-$24:43:02.55 & 4.94 & $G_{\rm BP}$ flux & 19.08$\pm$0.02 & 15.59$\pm$0.02 & 14.96$\pm$0.03 & -27.2 \\
        SPX J1044$-$4150 & 10:44:42.01 & $-$41:50:13.39 & 4.51 & fiducial & 18.95$\pm$0.02 & 15.82$\pm$0.02 & 15.51$\pm$0.04 & -26.4 \\
        SPX J1059$+$7738 & 10:59:51.14 & $+$77:38:13.70 & 5.51 & $G_{\rm RP}$ faint & 19.23$\pm$0.04 & 15.87$\pm$0.02 & 15.18$\pm$0.03 & -27.9 \\
        SPX J1150$-$2211 & 11:50:09.67 & $-$22:11:24.83 & 4.53 & fiducial & 19.17$\pm$0.02 & 15.18$\pm$0.02 & 14.78$\pm$0.03 & -26.6 \\
        SPX J1152$-$1805${^b}$ & 11:52:20.15 & $-$18:05:22.08 & 4.15 & fiducial & 18.17$\pm$0.04 & 13.68$\pm$0.01 & 13.20$\pm$0.01 & -28.0 \\
        \hline
    \end{tabular}
    \caption{New $z>4$ quasars discovered using SPHEREx spectrophotometry. Redshifts are determined via a Gaussian fit to the observed H$\alpha$ emission line. The $G_\text{RP}$ magnitudes are Vega magnitudes from the \textit{Gaia} DR3 catalog, while $W1$ and $W2$ are Vega magnitudes from the CatWISE2020 catalog.\\
             $^a$ Optical confirmation spectrum in ESO archive.\\
             $^b$ Optical confirmation spectrum in this work (see Section~\ref{sec:fup}).}
    \label{tab:hiz}
\end{table*}

\begin{table*}[]
    \ContinuedFloat
    \centering
    \begin{tabular}{lccclcccc}
        Name & RA & Dec & $z$ & Selection & $G_{\rm RP}$ & $W1$ & $W2$ & $M_{1450}$ \\
        \hline \hline 
        SPX J1309$-$7140 & 13:09:45.47 & $-$71:40:43.27 & 4.52 & fiducial & 19.12$\pm$0.04 & 13.82$\pm$0.02 & 13.38$\pm$0.01 & -28.2 \\
        SPX J1331$+$1511 & 13:31:34.15 & $+$15:11:15.32 & 4.32 & fiducial & 18.72$\pm$0.04 & 14.65$\pm$0.01 & 14.12$\pm$0.01 & -27.9 \\
        SPX J1333$+$8030 & 13:33:47.44 & $+$80:30:50.76 & 4.55 & fiducial & 18.74$\pm$0.04 & 15.64$\pm$0.02 & 15.18$\pm$0.02 & -27.3 \\
        SPX J1344$-$5123 & 13:44:20.80 & $-$51:23:53.23 & 4.87 & fiducial & 18.16$\pm$0.05 & 14.75$\pm$0.02 & 14.23$\pm$0.02 & -27.9 \\
        SPX J1412$-$4052 & 14:12:21.58 & $-$40:52:54.86 & 4.84 & fiducial & 18.87$\pm$0.02 & 15.67$\pm$0.02 & 15.29$\pm$0.03 & -27.7 \\
        SPX J1423$+$4413 & 14:23:16.04 & $+$44:13:11.57 & 4.12 & fiducial & 19.15$\pm$0.03 & 15.10$\pm$0.02 & 14.58$\pm$0.02 & -26.4 \\
        SPX J1444$-$7142 & 14:44:10.42 & $-$71:42:25.78 & 4.27 & fiducial & 18.32$\pm$0.02 & 14.95$\pm$0.02 & 14.57$\pm$0.02 & -27.2 \\
        SPX J1446$-$2750 & 14:46:19.57 & $-$27:50:41.66 & 4.37 & fiducial & 19.04$\pm$0.02 & 15.79$\pm$0.03 & 15.29$\pm$0.04 & -26.8 \\
        SPX J1526$-$7213 & 15:26:38.04 & $-$72:13:18.74 & 4.71 & fiducial & 19.12$\pm$0.05 & 14.70$\pm$0.01 & 14.29$\pm$0.01 & -26.7 \\
        SPX J1559$-$2126 & 15:59:50.82 & $-$21:26:13.88 & 4.37 & fiducial & 19.07$\pm$0.05 & 14.91$\pm$0.02 & 14.43$\pm$0.03 & -27.1 \\
        SPX J1605$-$3050 & 16:05:42.91 & $-$30:50:36.52 & 5.07 & fiducial & 18.92$\pm$0.03 & 14.64$\pm$0.02 & 14.05$\pm$0.02 & -27.5 \\
        SPX J1634$+$6734 & 16:34:43.92 & $+$67:34:28.94 & 4.70 & fiducial & 19.16$\pm$0.07 & 16.12$\pm$0.02 & 15.60$\pm$0.03 & -27.4 \\
        SPX J1651$+$1516 & 16:51:59.95 & $+$15:16:03.23 & 5.49 & fiducial & 19.11$\pm$0.01 & 15.93$\pm$0.03 & 15.59$\pm$0.04 & -28.2 \\
        SPX J1654$+$1102 & 16:54:42.46 & $+$11:02:32.46 & 4.59 & fiducial & 17.96$\pm$0.03 & 14.66$\pm$0.02 & 14.34$\pm$0.02 & -27.6 \\
        SPX J1701$-$1157 & 17:01:53.01 & $-$11:57:11.81 & 4.29 & fiducial & 19.04$\pm$0.07 & 15.65$\pm$0.02 & 15.30$\pm$0.04 & -26.9 \\
        SPX J1716$+$4638 & 17:16:10.47 & $+$46:38:58.83 & 4.45 & fiducial & 18.58$\pm$0.02 & 15.37$\pm$0.02 & 14.98$\pm$0.02 & -27.7 \\
        SPX J1726$-$1021 & 17:26:48.48 & $-$10:21:52.71 & 4.43 & fiducial & 18.66$\pm$0.03 & 14.31$\pm$0.01 & 13.93$\pm$0.01 & -26.9 \\
        SPX J1744$+$1044 & 17:44:46.90 & $+$10:44:57.32 & 4.72 & fiducial & 18.91$\pm$0.03 & 15.97$\pm$0.03 & 15.43$\pm$0.04 & -26.7 \\
        SPX J1758$+$2659 & 17:58:59.05 & $+$26:59:57.86 & 4.62 & fiducial & 18.28$\pm$0.04 & 14.76$\pm$0.02 & 14.37$\pm$0.02 & -28.7 \\
        SPX J1835$+$4334 & 18:35:56.29 & $+$43:34:53.37 & 5.12 & fiducial & 18.69$\pm$0.03 & 15.43$\pm$0.02 & 15.04$\pm$0.02 & -28.3 \\
        SPX J1905$+$5655 & 19:05:35.27 & $+$56:55:36.98 & 4.43 & fiducial & 18.57$\pm$0.03 & 14.93$\pm$0.01 & 14.51$\pm$0.01 & -27.6 \\
        SPX J1910$+$3500 & 19:10:11.84 & $+$35:00:08.43 & 5.19 & $G_{\rm RP}$ faint & 19.37$\pm$0.02 & 15.73$\pm$0.03 & 15.11$\pm$0.03 & -27.7 \\
        SPX J1912$+$4657 & 19:12:30.73 & $+$46:57:44.86 & 4.44 & fiducial & 17.67$\pm$0.03 & 14.70$\pm$0.01 & 14.37$\pm$0.01 & -28.0 \\
        SPX J1927$+$5200 & 19:27:46.29 & $+$52:00:04.56 & 4.37 & fiducial & 17.73$\pm$0.05 & 14.32$\pm$0.01 & 13.88$\pm$0.01 & -28.1 \\
        SPX J1946$-$6015 & 19:46:52.96 & $-$60:15:57.16 & 4.79 & fiducial & 18.95$\pm$0.06 & 15.64$\pm$0.03 & 15.07$\pm$0.03 & -27.9 \\
        SPX J1956$-$2617 & 19:56:13.99 & $-$26:17:28.76 & 4.58 & fiducial & 18.86$\pm$0.02 & 15.50$\pm$0.02 & 15.15$\pm$0.04 & -27.3 \\
        SPX J1959$+$0333 & 19:59:12.82 & $+$03:33:53.99 & 4.20 & fiducial & 18.72$\pm$0.02 & 14.33$\pm$0.01 & 14.03$\pm$0.02 & -27.5 \\
        SPX J2003$-$6049 & 20:03:35.52 & $-$60:49:01.76 & 4.86 & $G_{\rm RP}$ faint & 19.16$\pm$0.01 & 15.83$\pm$0.03 & 15.13$\pm$0.03 & -28.0 \\
        SPX J2018$-$4539 & 20:18:59.38 & $-$45:39:24.60 & 5.44 & $G_{\rm BP}$ flux & 19.58$\pm$0.05 & 16.23$\pm$0.03 & 15.40$\pm$0.04 & -26.9 \\
        SPX J2040$+$6741${^b}$ & 20:40:55.75 & $+$67:41:12.64 & 4.68 & fiducial & 18.61$\pm$0.03 & 14.90$\pm$0.01 & 14.46$\pm$0.01 & -28.1 \\
        SPX J2047$-$3935 & 20:47:27.81 & $-$39:35:58.25 & 4.69 & fiducial & 19.16$\pm$0.02 & 15.55$\pm$0.02 & 15.14$\pm$0.03 & -27.4 \\
        SPX J2051$+$2156${^b}$ & 20:51:34.59 & $+$21:56:36.18 & 5.31 & $G_{\rm RP}$ faint & 19.03$\pm$0.03 & 15.32$\pm$0.02 & 14.48$\pm$0.02 & -27.6 \\
        SPX J2103$+$1838 & 21:03:16.91 & $+$18:38:03.93 & 4.62 & fiducial & 18.07$\pm$0.03 & 15.21$\pm$0.02 & 14.80$\pm$0.02 & -28.1 \\
        SPX J2109$+$3003${^b}$ & 21:09:25.28 & $+$30:03:56.37 & 4.20 & fiducial & 17.66$\pm$0.05 & 13.56$\pm$0.01 & 13.09$\pm$0.01 & -28.7 \\
        SPX J2147$+$2357 & 21:47:45.62 & $+$23:57:04.50 & 4.80 & fiducial & 19.14$\pm$0.03 & 15.40$\pm$0.03 & 14.86$\pm$0.03 & -28.0 \\
        SPX J2148$-$2210 & 21:48:19.36 & $-$22:10:53.73 & 4.49 & fiducial & 18.64$\pm$0.01 & 15.46$\pm$0.02 & 15.04$\pm$0.03 & -28.3 \\
        SPX J2237$-$3321 & 22:37:14.18 & $-$33:21:47.23 & 4.46 & fiducial & 19.04$\pm$0.03 & 15.80$\pm$0.03 & 15.36$\pm$0.04 & -26.5 \\
        SPX J2238$-$4402${^a}$ & 22:38:14.89 & $-$44:02:59.13 & 4.49 & fiducial & 18.78$\pm$0.03 & 16.00$\pm$0.03 & 15.53$\pm$0.04 & -27.6 \\
        SPX J2238$+$4140 & 22:38:44.22 & $+$41:40:19.60 & 5.73 & fiducial & 19.20$\pm$0.02 & 15.33$\pm$0.02 & 14.53$\pm$0.02 & -27.9 \\
        SPX J2303$+$3456${^b}$ & 23:03:08.65 & $+$34:56:20.77 & 5.42 & fiducial & 18.49$\pm$0.03 & 14.39$\pm$0.01 & 13.60$\pm$0.01 & -29.0 \\
        SPX J2303$-$4156 & 23:03:13.53 & $-$41:56:29.32 & 4.62 & fiducial & 19.16$\pm$0.02 & 15.92$\pm$0.03 & 15.52$\pm$0.04 & -27.2 \\
        SPX J2324$-$4801${^a}$ & 23:24:43.38 & $-$48:01:48.90 & 5.13 & $G_{\rm RP}$ faint & 19.31$\pm$0.01 & 16.11$\pm$0.03 & 15.56$\pm$0.04 & -27.0 \\

        \hline
    \end{tabular}
    \caption{Continued.}
    \label{tab:hiz2}
\end{table*}

\section{Newly discovered $z<4$ quasars}

In Table~\ref{tab:lowz} we provide the list of 247 new $z<4$ quasars from our fiducial selection confirmed by SPHEREx spectrophotometry. Numerous additional $z<4$ quasars were also selected by the $G_{\rm RP}$ faint and $G_{\rm BP}$ flux selections, however, as these selections were tuned specifically for high redshift discovery, we did not record them.

\begin{figure*}
    \centering
    \includegraphics[width=0.9\linewidth]{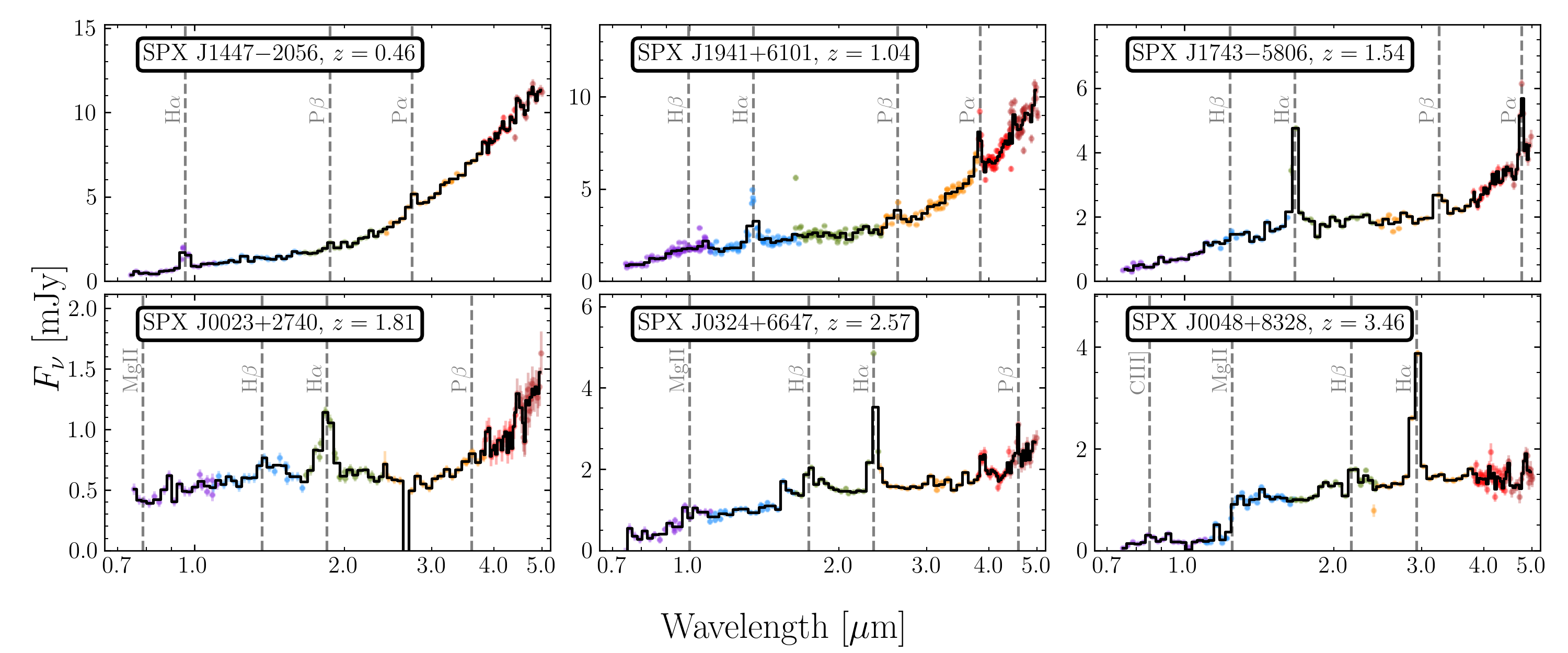}
    \caption{Example SPHEREx spectrophotometry of newly discovered $z<4$ quasars.}
    \label{fig:lowz}
\end{figure*}

\begin{figure*}
    \centering
    \includegraphics[width=0.32\linewidth]{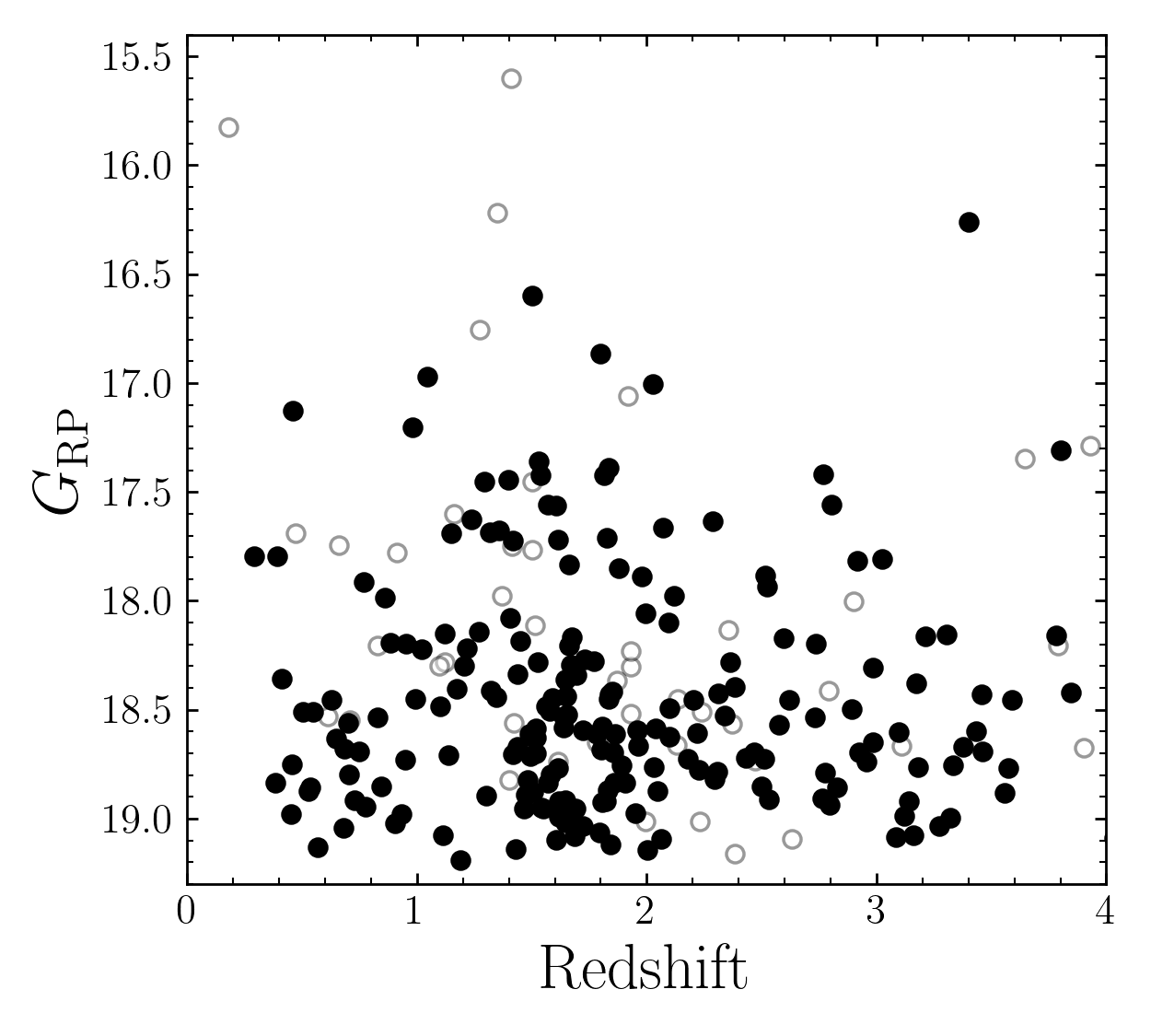}
    \includegraphics[width=0.32\linewidth]{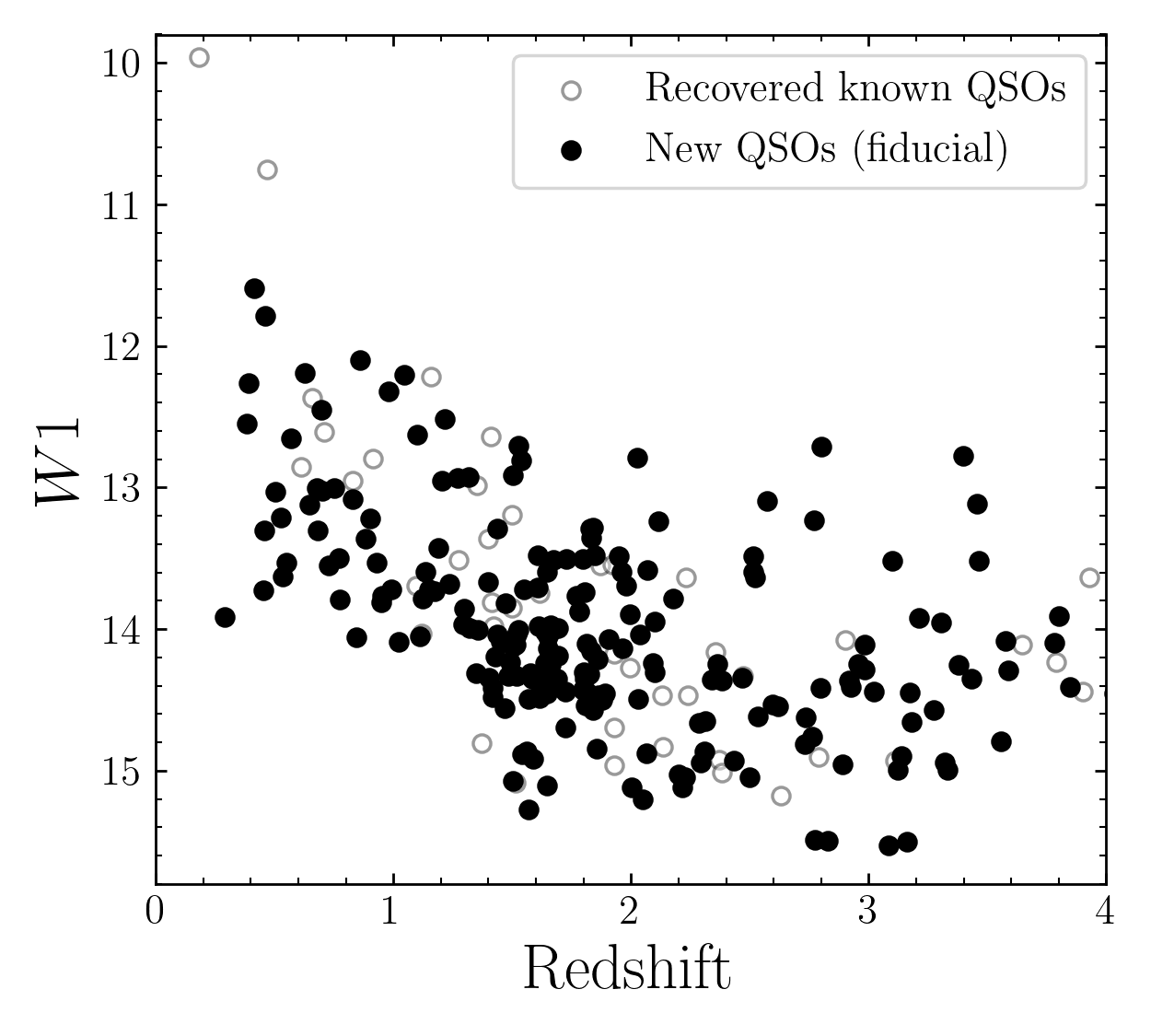}
    \includegraphics[width=0.32\linewidth]{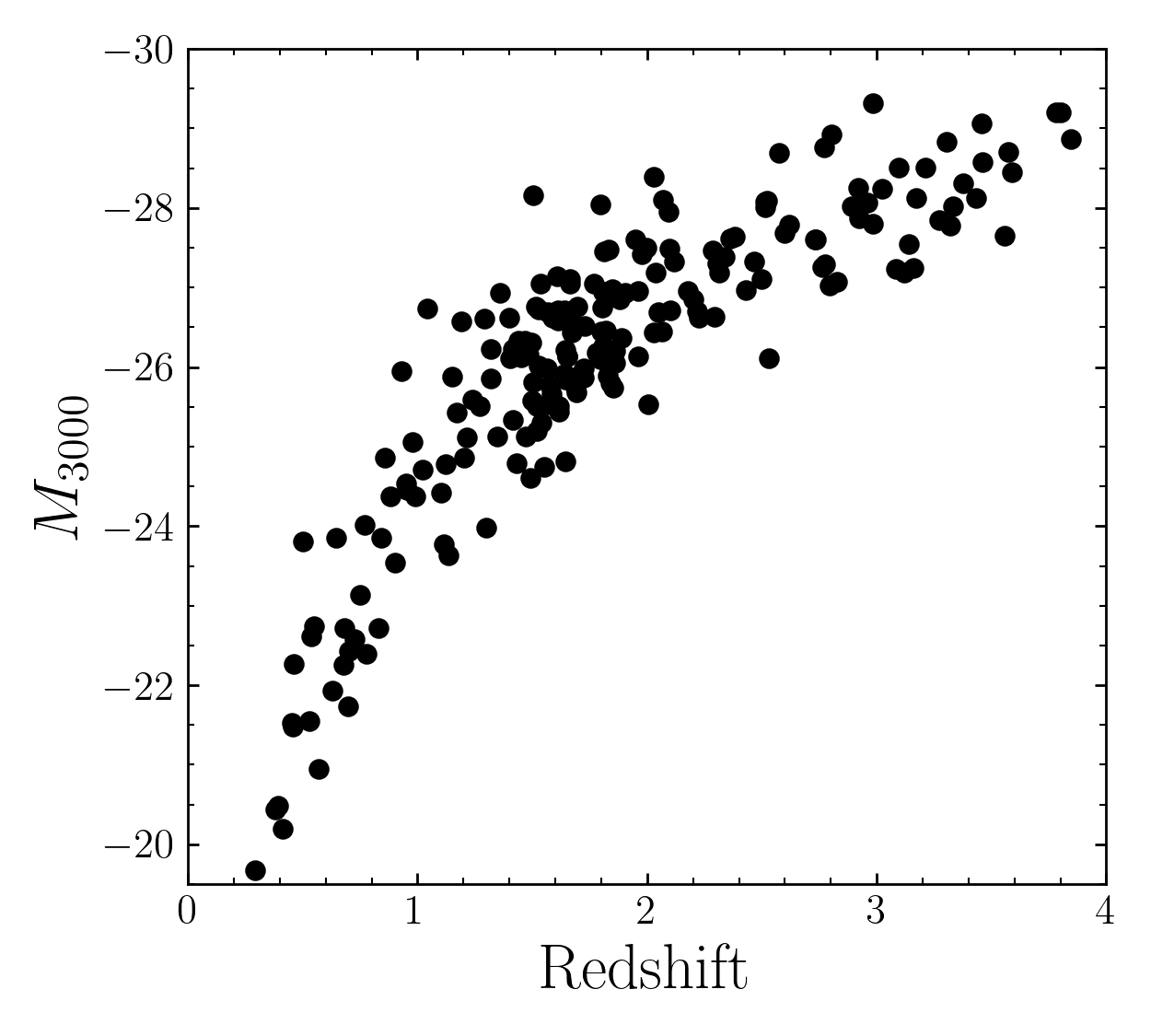}
    \caption{Distribution of $G_{\rm RP}$ magnitude (left), $W1$ magnitude (middle), and $M_{3000}$ from template fitting (right) for the newly discovered (solid points) and recovered known (open points) $z<4$ quasars.}
    \label{fig:rp_w1_lowz}
\end{figure*}

\begin{table*}[]
    \centering
    \begin{tabular}{lccccccc}
        Name & RA & Dec & $z$ & $G_{\rm RP}$ & $W1$ & $W2$ & $M_{3000}$ \\
        \hline \hline 
        SPX J0009$-$3503 & 00:09:11.49 & $-$35:03:38.32 & 2.73 & 18.53$\pm$0.03 & 14.81$\pm$0.02 & 14.11$\pm$0.02 & -27.6 \\
        SPX J0010$-$3936 & 00:10:34.72 & $-$39:36:07.09 & 1.24 & 17.63$\pm$0.03 & 13.68$\pm$0.01 & 12.40$\pm$0.01 & -25.6 \\
        SPX J0023$+$2740 & 00:23:29.43 & $+$27:40:57.42 & 1.81 & 17.42$\pm$0.06 & 14.11$\pm$0.01 & 12.99$\pm$0.01 & -27.5 \\
        SPX J0035$-$5041 & 00:35:09.23 & $-$50:41:08.12 & 0.78 & 18.94$\pm$0.02 & 13.79$\pm$0.01 & 12.41$\pm$0.01 & -22.4 \\
        SPX J0036$-$5408 & 00:36:54.01 & $-$54:08:43.98 & 2.18 & 18.73$\pm$0.03 & 13.78$\pm$0.01 & 12.52$\pm$0.01 & -27.0 \\
        SPX J0047$-$8941 & 00:47:11.84 & $-$89:41:57.46 & 0.99 & 18.45$\pm$0.03 & 13.72$\pm$0.02 & 12.49$\pm$0.01 & -24.4 \\
        SPX J0048$+$8328 & 00:48:29.91 & $+$83:28:00.93 & 3.46 & 18.43$\pm$0.06 & 13.11$\pm$0.01 & 12.68$\pm$0.01 & -29.1 \\
        SPX J0050$+$4318 & 00:50:28.88 & $+$43:18:54.84 & 3.17 & 18.38$\pm$0.05 & 14.45$\pm$0.02 & 14.00$\pm$0.01 & -28.1 \\
        SPX J0053$-$0242 & 00:53:27.05 & $-$02:42:16.85 & 1.42 & 18.70$\pm$0.02 & 14.48$\pm$0.01 & 13.08$\pm$0.01 & -25.3 \\
        SPX J0100$-$4211 & 01:00:08.92 & $-$42:11:41.10 & 1.53 & 17.36$\pm$0.05 & 12.71$\pm$0.01 & 11.20$\pm$0.01 & -26.7 \\
        SPX J0101$-$3426 & 01:01:42.59 & $-$34:26:33.72 & 1.65 & 18.52$\pm$0.03 & 14.05$\pm$0.01 & 12.68$\pm$0.01 & -26.1 \\
        SPX J0117$+$2223 & 01:17:55.29 & $+$22:23:18.35 & 1.47 & 18.89$\pm$0.03 & 13.81$\pm$0.01 & 12.35$\pm$0.01 & -25.1 \\
        SPX J0118$+$1620 & 01:18:49.51 & $+$16:20:14.93 & 1.72 & 19.03$\pm$0.01 & 14.69$\pm$0.02 & 13.22$\pm$0.01 & -25.9 \\
        SPX J0126$+$4600 & 01:26:32.91 & $+$46:00:11.85 & 1.62 & 18.92$\pm$0.03 & 14.36$\pm$0.01 & 13.01$\pm$0.01 & -25.4 \\
        SPX J0134$-$5401 & 01:34:32.94 & $-$54:01:36.57 & 1.83 & 17.39$\pm$0.12 & 13.36$\pm$0.01 & 12.39$\pm$0.01 & -27.5 \\
        SPX J0150$+$4659 & 01:50:08.13 & $+$46:59:40.83 & 1.17 & 18.40$\pm$0.06 & 13.73$\pm$0.01 & 12.52$\pm$0.01 & -25.4 \\
        SPX J0153$+$0851 & 01:53:16.60 & $+$08:51:21.60 & 3.85 & 18.42$\pm$0.03 & 14.41$\pm$0.01 & 13.91$\pm$0.01 & -28.9 \\

        ...
    \end{tabular}
    \caption{New $z<4$ quasars discovered using SPHEREx spectrophotometry. The full table containing \lowznew objects can be found in the online journal.}
    \label{tab:lowz}
\end{table*}
\end{appendix}

\end{document}